\documentclass[twocolumn,showpacs, aps, superscriptaddress, eqsecnum, prd, 
notitlepage, showkeys, nofootinbib, floatfix]{revtex4-2}

\usepackage{graphicx}
\usepackage{dcolumn}
\usepackage{bm}

\usepackage{amssymb}
\usepackage{amsmath}
\usepackage{graphicx}
\usepackage{dcolumn}
\usepackage[pdfencoding=auto]{hyperref}
\usepackage{color,units}
\usepackage[dvipsnames]{xcolor} 
\usepackage{aas_macros}
\usepackage{xspace}
\usepackage[normalem]{ulem} 
\usepackage{float}
\usepackage{orcidlink}
\usepackage{multirow}
\usepackage{booktabs}
\newcommand{\bilby}{\texttt{Bilby}}
\newcommand{\dynesty}{\texttt{Dynesty}}

\newcommand{\gwMay}{\texttt{GW190521}}
\newcommand{\gwSep}{\texttt{GW190929}}
\newcommand{\gwFirst}{\texttt{GW150914}}
\newcommand{\asd}[1]{\text{ASD}}
\newcommand{\tpowerbilby}{\texttt{\textbf{tPowerBilby}}}

\newcommand{\giturl}{\href{https://github.com/NirGutt/tPowerBilby}{\tpowerbilby}}

\begin{document}

\title{Modelling noise in gravitational-wave observatories with transdimensional models}

\author{Nir Guttman}
\email{nir.guttman@monash.edu}
\affiliation{School of Physics and Astronomy, Monash University, VIC 3800, Australia}
\affiliation{OzGrav: The ARC Centre of Excellence for Gravitational-Wave Discovery, Clayton, VIC 3800, Australia}

\author{Paul D. Lasky \orcidlink{0000-0003-3763-1386}}
\affiliation{School of Physics and Astronomy, Monash University, VIC 3800, Australia}
\affiliation{OzGrav: The ARC Centre of Excellence for Gravitational-Wave Discovery, Clayton, VIC 3800, Australia}

\author{Eric Thrane \orcidlink{0000-0002-4418-3895}}
\affiliation{School of Physics and Astronomy, Monash University, VIC 3800, Australia}
\affiliation{OzGrav: The ARC Centre of Excellence for Gravitational-Wave Discovery, Clayton, VIC 3800, Australia}

\begin{abstract}
Modelling noise in gravitational-wave observatories is crucial for accurately inferring the properties of gravitational-wave sources. We introduce a transdimensional Bayesian approach to characterise the noise in ground-based gravitational-wave observatories using the Bayesian inference software \texttt{Bilby}. The algorithm models broadband noise with a combination of power laws; narrowband features with Lorentzians; and shapelets to capture any additional features in the data. 
We show that our noise model provides a significantly improved fit of the LIGO and Virgo noise amplitude spectral densities compared to currently available noise fits obtained with on-source data segments.
We perform astrophysical inference on well-known events in the third Gravitational-Wave Transient Catalog using our noise model and
observe shifts of up to $7\%$ in the $90\%$ boundaries of credible intervals for some parameters.
We discuss plans to deploy this framework systematically for gravitational-wave inference along with possible areas of improvement.
\end{abstract}

\maketitle

\section{Introduction}
Due to the complex nature of terrestrial gravitational-wave interferometers, which consist of numerous subsystems, each with their own noise characteristics, it is probably impossible to develop a comprehensive noise model from first principles~\citep{ligo_det,virgo_det,Ligo_data_analysis_guide}. 
That said, a number of noise sources can be modelled.
Known noise sources include seismic noise at low frequencies (\(f \lesssim 30 \text{ Hz}\)), thermal noise at mid-range frequencies (\(30 \lesssim f \lesssim 100 \text{ Hz}\)), and quantum noise at high frequencies (\(f \gtrsim 100 \text{ Hz}\)). These three noise sources are regarded as broadband noise sources.
In addition, there are known narrowband noise features present resulting from several sources such as power lines at $\unit[60]{Hz}$ in the USA ($\unit[50]{Hz}$ in Europe) and their harmonics, suspension wire resonance, and calibration lines that are intentionally added. 
Finally, there are other features that do not resemble the slowly-varying broadband noise sources or the narrowband lines. 
For example, there are features in the noise curve around 30\,Hz and another at 500\,Hz that are not shaped like narrowband lines, but which are not shaped like slowly varying broadband noise either.

Gravitational-wave interferometers produce a strain time series $h(t)$ which is sampled and digitized at $\unit[16384]{Hz}$. 
As our starting point, we assume that the strain noise is approximately stationary and Gaussian. It is best characterized in the frequency domain, where the distribution of strain noise in an interferometer $n(f)$ is characterised by the (single-sided) power spectral density [PSD; $P(f)$], defined as the variance of the frequency-domain strain noise:\footnote{In reality, the noise properties may deviate from these assumptions. For example, our ability to describe the noise with a diagonal matrix in the frequency domain arises from the assumption that the noise in each frequency bin is uncorrelated. This is not exactly true in practice, and the resulting error can produce systematic errors comparable to the effects we study here \citep{windows}.}
\begin{align}\label{eq:PSD}
    \langle n^*(f) n(f') \rangle = \frac{1}{2} \delta(f-f') \, P(f) .
\end{align}

\begin{figure}
    \centering
    \includegraphics[width=\columnwidth]{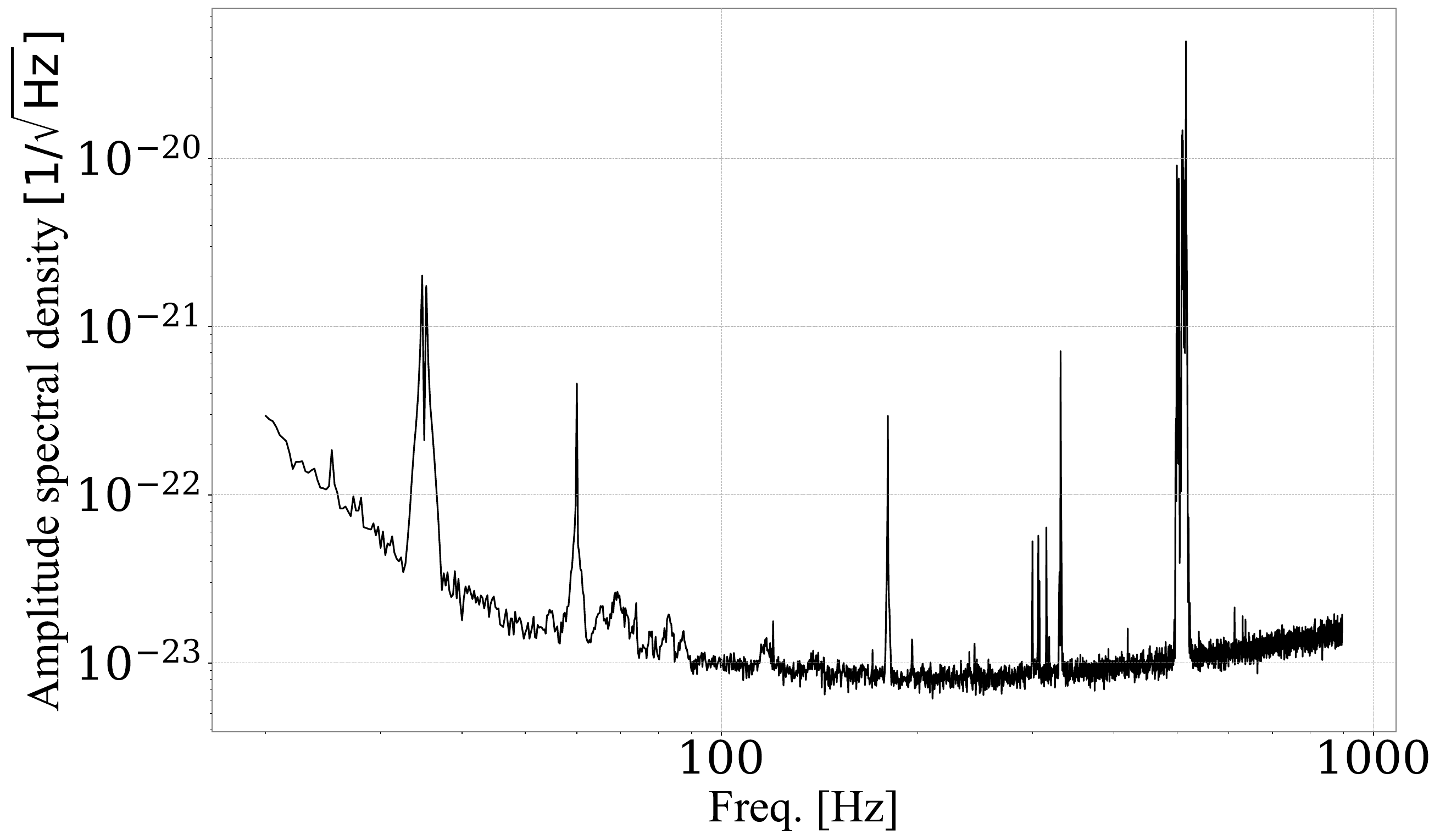}
    \caption{A typical noise curve for the LIGO Livingston observatory, showing the strain amplitude spectral density, obtained using Welch's method.}  
    \label{fig:GW150914_plain_asd}
\end{figure}

Figure~\ref{fig:GW150914_plain_asd} presents an example of a typical amplitude spectral density curve
\begin{align}
\sigma(f) = \sqrt{P(f)} ,
\end{align}
using data adjacent to the first ever detected gravitational wave signal \gwFirst{}~\cite{GW150914}. 
In this plot, $\sigma(f)$ is calculated with Welch's method~\citep{welch}, which we describe below.

In order to infer the properties of gravitational-wave signals from various sources such as binary black holes, gravitational-wave astronomers employ Bayesian inference, which relies on an accurate description of the noise. 
Consequently, accurate noise models are essential for valid astrophysical conclusions. 

In this study, we introduce an easy-to-use algorithm for gravitational-wave interferometer noise estimation based on Bayesian transdimensional sampling, leveraging theoretical considerations alongside data-driven methods. 
Our method demonstrates improvements over other noise estimation techniques in three illustrative examples. 
Additionally, we evaluate the impact of uncertainties in the noise model on astrophysical parameters for these three examples.

As part of this work, we provide an open-source software package \tpowerbilby{}\footnote{https://github.com/NirGutt/tPowerBilby} that allows users to estimate the gravitational-wave detector noise using publicly-available data from the Gravitational-Wave Open Science Centre~\cite[GWOSC;][]{GWOSC1,GWOSC2}. 
The package is built around the transdimensional extension~\cite{tbilby} to the widely used \bilby{}~\cite{bilby,romeroshaw20} package.
We describe software details in Appendix~\ref{appx:tPowerbilby}.

Various methods have been employed in order to estimate the amplitude spectral density $\sigma(f)$.
The most common approach is to estimate it empirically using off-source data with an estimator of the form:
\begin{align}
    \widehat\sigma(f_k)^2 = \frac{1}{2} \overline{n(f_k)^2} ,
\end{align}
where the overline denotes either a median (or average) over $N \approx 32$ ``off-source'' segments, each of which typically corresponds to \( T =4\) seconds of sampled strain data, in the vicinity of a gravitational-wave event.\footnote{If one uses the median, one has to include an additional bias correction factor~\cite{median_factor}.}
Here, \( n(f_k) \) is the discrete Fourier transform of \( n(t) \), the noise time series, at frequency bin $k$. Practically, the Fourier transform is computed from a finite time series of length \( T \). This process is repeated multiple times, and typically the average or the median is taken using Welch's method~\cite{welch}.

This \textit{empirical} approach has the advantage that it makes minimum assumptions---only that the off-source data has similar noise properties to the ``on-source'' data containing the gravitational-wave signal.
The disadvantage of the empirical approach is that there are reasonable assumptions we can make that produce a more accurate noise estimate; namely, that the noise varies smoothly as a function of frequency except for occasional narrowband features.

\texttt{BayesLine}~\cite{Littenberg_2015,BayesWaveUpdated2024} is widely used software that estimates the noise PSD of gravitational-wave interferometers using transdimensional Bayesian inference. To overcome the limitations of empirical models, \texttt{BayesLine} models the smoothly varying features of the PSD with spline curves. 
Our algorithm draws inspiration from \texttt{BayesLine}.
However, it is implemented as part of \bilby{} in order to leverage \bilby's extensive user base and flexibility for future extensions. 
We discuss further details regarding the differences between the methods in Section~\ref{sec:conclusions}.

The remainder of this paper is organized as follows. 
In Section~\ref{sec:bayesian_framework}, we lay out the problem of noise estimation in the context of Bayesian inference and describe the details of our noise model.
In Section~\ref{sec:study_of_real_data}, we apply this model to data from GWTC-3 in order to ascertain the goodness of fit.
We show how measurements of astrophysical parameters are affected by comparing posterior distributions of these parameters obtained using our noise model and through other means.
In Section~\ref{sec:conclusions} we discuss the implications of these results and discuss plans for future work.

\section{Noise model}\label{sec:bayesian_framework}
\subsection{Overview}

Assuming Gaussian, stationary noise, the strain data are distributed according to the Whittle likelihood \citep{Thrane_2019}:
\begin{equation}\label{eq:likelihood}
\mathcal{L}\left( \vec{d} | \theta ,\Lambda \right) = 
\prod_k
\frac{1}{2 \pi \sigma_k^2\left( \Lambda \right)}\exp\left(-2 \Delta f \frac{\left|d_k - \mu_k(\theta) \right|^{2}}{\sigma_k^2\left(\Lambda \right)}\right).
\end{equation}
Here, $d_k$ represents the strain in frequency bin $k$.
The data consists of signal $s$ and noise $n$ so that 
\begin{align}
    d_k = s_k + n_k .
\end{align}
The variable $\mu_k(\theta)$ represents the predicted astrophysical strain in a frequency bin $k$, which depends on parameters $\theta$, e.g., the masses of merging black holes. 
Meanwhile, $\Delta f$ is the width of the frequency bin and $\sigma_k$
is the noise amplitude spectral density.
Since we are modelling the noise, $\sigma_k$ also depends on parameters $\Lambda$.

Our noise model consists of two parts: broadband noise $\sigma_{\rm BB}(f)$ and narrowband noise $\sigma_{\rm NB}(f)$. Each part is a superposition of two types of functions.
The broadband noise is described by (1) power laws $\sigma_{\rm PL}(f)$ as in~\cite{Cahillane_2022}, and noise that is not well described by the power laws is captured using (2) shapelet functions~\cite{Shapelets} $\sigma_{\rm BBS}(f)$.
The narrowband noise is modeled using (1) Lorentzian functions, following Ref.~\citep{Littenberg_2015}, denoted $\sigma_{\rm line}(f)$. Similar to the broadband part, additional narrowband features in the data that are not well described by the Lorentzian functions are modeled using (2) shapelet functions $\sigma_{\rm NBS}(f)$.

Each noise component described above can have multiple pieces so that, for example, the broadband noise might include $N_\text{PL}=3$ power-law components and
$N_\text{BBS}=2$ broadband shapelets, while the narrowband might include $N_\text{line}=10$ and $N_\text{NBS}=1$ narrowband features. 
Our model for the amplitude spectral density is:
\begin{align}\label{eq:main_fit}
    \sigma(f,\Lambda) =& \max \Bigg[  \sigma_{\text{BB}}(f,\Lambda_\text{BB}) , \sigma_{\text{NB}}(f,\Lambda_\text{NB}) \Bigg],
\end{align}
where
\begin{align}\label{eq:sigma_BB_NB}
    \sigma_{\text{BB}}(f,\Lambda_\text{BB}) &= \sum_{i}^{N_{\text{PL}}} \sigma_\text{PL}^{i}(f,\Lambda_{\text{PL}}^{i}) + \sum_{l}^{N_{\text{BBS}}}\sigma_{\text{BBS}}^{l}(f,\Lambda_{\text{BBS}}^{l}),\\ \nonumber
    \sigma_{\text{NB}}(f,\Lambda_\text{NB})&=\sum_{k}^{N_{\text{line}}} \sigma_\text{line}^{k}(f,\Lambda_{\text{line}}^{k})+
\sum_{j}^{N_{\text{NBS}}}\sigma_{\text{NBS}}^{j}(f,\Lambda_{\text{NBS}}^{j}).   
\end{align}

The terms $(\sigma_{\text{PL}}, \sigma_\text{BBS}, \sigma_\text{line}, \sigma_\text{NBS})$ are ``component functions'' for each type of noise (defined below).
They are parameterised by dedicated sets of parameters $( \Lambda_\text{PL}, \Lambda_\text{BBS}, \Lambda_\text{line}, \Lambda_\text{NBS}$).
For each frequency bin, $\sigma(f)$ is determined by either the broadband model $\sigma_\text{BB}(f)$ or the narrowband model $\sigma_\text{NB}(f)$---whichever is larger. 
Taking the maximum helps to decouple the broadband and narrowband fits, which in turn reduces the complexity of the problem; see Sec.~\ref{subsec:sampling}.  
The maximum operation becomes relevant only in the frequency bins identified by the algorithm described in Appendix~\ref{appx:preprocessing}. 
In the following subsections, we provide a detailed description of each of the noise type and its parameters. Table~\ref{tab:model_components} summarizes the model parameters.

\begin{table}
    \centering
    \begin{tabular}{ccc}
    \hline
        Component & Parameter Name & Description \\
    \hline
        \multirow{4}{*}{\shortstack{Broadband \\ \\ $\sigma_\text{PL}$}} 
        & $N_\text{PL}$ [-] & Number of functions \\
        & $\alpha$ [-] & Spectral index \\
        & $A_{\text{PL}}$ [$\mathrm{Hz}^{-\alpha -1/2}$] & Amplitude \\
    \hline
        \multirow{7}{*}{\shortstack{Narrowband \\ \\ $\sigma_\text{line}$}} 
        & $N_\text{line}$ [-] & Number of functions \\
        & $\Gamma$ [Hz] & Width \\
        & $\zeta$ [-] & Damping start width \\
        & $\tau$ [$\mathrm{Hz}^{-1}$] & Damping rate  \\
        & $f_{l}$ [Hz] & Peak frequency \\
        & $A_{l}$ [$\mathrm{Hz}^{-1/2}$] & Amplitude \\
    \hline
        \multirow{6}{*}{\shortstack{Shapelets \\ \\ $\sigma_\text{BBS/NBS}$}} 
        & $N_\text{BBS/NBS}$ [-] & Number of functions \\
        & $\text{deg}_\text{BBS/NBS}$ [-] & Maximum shapelet degree \\
        & $\beta_\text{BBS/NBS}$ [Hz] & Width \\
        & $f_\text{sh}^\text{BBS/NBS}$ [Hz] & Peak frequency \\
        & $A_\text{sh}^\text{BBS/NBS}$ [-] & Amplitude \\
    \hline
    \end{tabular}
    \caption{Components of the model and their associated parameters. BBS and NBS refer to the broadband and narrowband shapelets, respectively, and should be considered as distinct sets of parameters.}
    \label{tab:model_components}
\end{table}

\subsection{Broadband Power Law Function}
For the broadband noise component, we employ a sum of $N_{\text{PL}}$ power law functions, each function is  described by
\begin{equation}
    \sigma_{\text{PL}}(f, A_{\text{PL}}, \alpha) = A_{\text{PL}}f^\alpha,\label{eq:power_laws}
\end{equation}
where \( A_{\text{PL}} \) denotes the amplitude and \( \alpha \) the spectral index.

\subsection{Narrowband Lorentzian Function}
The narrowband spectral lines are modeled with a Lorentzian function (as expected for a damped driven harmonic oscillator). 
To enhance flexibility and ensure a realistic description, we incorporate an exponential decaying tail:

\begin{equation}
    \sigma_\text{line}(f, \Gamma, f_{l}, A_{l}, \zeta, \tau) = e^{-\gamma(f)\left(|f-f_{l}| - \Gamma\zeta\right)} \frac{A_l \Gamma^2}{\Gamma^2 + (f-f_{l})^2},
\end{equation}

where $f_l$ is the central frequency, $A_l$ is the amplitude, and $\Gamma$ is the width. 
Meanwhile, \( \gamma(f) \) dictates the exponential decay to prevent long tails unsupported by data:
\[
\gamma(f) =
\begin{cases}
0 & \text{if } |f - f_{l}| < \Gamma \zeta \\
\tau & \text{if } |f - f_{l}| > \Gamma \zeta
\end{cases}.
\]
Here, $\tau$ is the characteristic decay scale, and \( \zeta \) represents the distance in units of the Lorentzian width from the central frequency \( f_{l} \).
Finally, the narrowband noise is represented as a sum of $N_{\text{line}}$ such lines.

\subsection{Shapelet Functions}
To accommodate noise components that exhibit arbitrary shapes in the data, we introduce shapelets---functions often used for their flexibility. Each shapelet is defined as:
\begin{equation}
\begin{split}
    S_{m}(f,f_\text{sh},A_\text{sh}^{m},\beta) = A_\text{sh}^{m} \cdot \beta^{-\frac{1}{2}} \left( 2^{m} \pi^{\frac{1}{2}} m! \right)^{-\frac{1}{2}}\times \\
    H_{m} \left( \beta^{-1} \left(f-f_\text{sh}\right) \right) e^{-\frac{1}{2} \left( \beta^{-1} \left(f-f_\text{sh}\right) \right)^2},
    \end{split}
\end{equation}
where \( m \) is the degree of the shapelet, \( A_\text{sh}^{m} \) is its amplitude of degree \( m \), \( \beta \) represents its typical width, \( f_\text{sh} \) is its central frequency, and \( H_m \) denotes the Hermite polynomial of order \( m \).
The noise component is defined as a sum of shapelets up to a certain degree, 
\begin{equation}\label{eq:shaplet}
\sigma_{\text{BBS/NBS}}(f,f_\text{sh},A_\text{sh},\beta) = \sum_{m}^{\text{deg}} S_{m}(f,f_\text{sh},A_\text{sh}^{m},\beta)
\end{equation}
These ``other'' broadband and narrowband noises are modeled as a superposition of $N_{\text{BBS/NBS}}$ instances of Eq.~\ref{eq:shaplet}. For clarity, the superscripts of the parameters BBS and NBS are omitted from Eq.~\ref{eq:shaplet} (due to the presence of the degree $m$ superscript); however, the parameters should be interpreted as detailed in Table~\ref{tab:model_components}.  

\subsection{Priors}\label{subsec:priors}
In the following subsections we provide a detailed description of each of the noise model parameters priors. Table~\ref{tab:prior} summarizes the
model parameter's priors.

\subsubsection{Broadband power law noise prior }
For the broadband noise, we utilise $N_{\text{PL}}$ power law functions, each consisting of two parameters: an amplitude $A_\text{PL}$ and a spectral index $\alpha$. The prior on $N_{\text{PL}}$ follows a discrete uniform distribution $\mathcal{DU}[0,5]$. 
The prior on the amplitude follows a log-uniform distribution  $\log \mathcal{U}[10^{-30}$,$10^{-13}]$,\footnote{For succinctness we frequently do not state units, but throughout units for each quantity are those presented in Table~\ref{tab:model_components}.} while the prior on the spectral index follows a conditional uniform distribution $\mathcal{CU}[-10,2]$, enforcing $\alpha_{i} > \alpha_{i-1}$. This means that as we add additional power law functions to describe the broadband noise, each new $\alpha$ must be greater than the previous one, effectively ordering the power law functions by increasing spectral indices. As a result, the prior volume contracts as more power-law functions are added, reducing the Occam penalty for introducing a more complex model compared to the previously added power-law functions. To mitigate this effect, we dynamically increase the maximum allowed prior value, ensuring an identical prior volume for each power-law function.
This ordering of spectral indices aids in distinguishing different contributions and ensures interpretability of this part of the model. Careful consideration is necessary when implementing this ordering scheme, as it may overly constrain the sampler, potentially limiting exploration of the entire prior space. Therefore, we selectively apply this solely to spectral indices, rather than to other model components. 

\subsubsection{Narrowband Lorentzian noise prior}
The narrowband noise description consists of $N_{\text{line}}$ Lorentzian functions with damped tails, each consisting of five parameters $\Gamma$, $\tau$, $\zeta$, $f_l$, and $A_l$. 
The prior on $N_{\text{line}}$ follows a discrete uniform distribution $\mathcal{DU}[0,20]$.  
The prior on the width parameter $\Gamma$ is a log-uniform distribution $\log \mathcal{U}[10^{-3},1]$. The priors on the tail damping parameter $\zeta$ is a truncated Gaussian, ranging from 0.1 to 5, centered at 2.7 with width of 1.1,
\begin{align}
    \mathcal{G}_t(\mu=2.7,\sigma=1.1),
\end{align}
while we keep $\tau$ fixed to 5.2. These values are determined by experimentation and found to be consistent between different interferometers and different time segments. 
The prior distribution of line locations $f_l$ is approximated by interpolation, covering the entire frequency range: $\texttt{Interp}[f_\text{min}, f_\text{max}]$, and is determined empirically by a line-finding algorithm described in Appendix~\ref{appx:lines_prior}, effectively setting a tight prior around the lines found in adjacent data. Finally, for the prior of the amplitude $A_l$, we introduce a conditional log-uniform distribution $\log \mathcal{CU}[A_l^\text{min}(f_l),A_l^\text{max}(f_l)]$. This is conditional on the parameter $f_l$ in the sense that the maximal and minimal value of the distribution are determined by $f_l$. We provide a detailed description of how the maximum and minimum values vary with frequency in Appendix~\ref{appx:lines_prior}.

\subsubsection{Shapelet noise prior}
The ``other'' noise is a superposition of $N_{\text{BBS/NBS}}$ components, where each component is a sum of shapelets. Each shaplelet is characterized by three parameters: $\beta$, $f_{\text{sh}}$, and $A_{\text{sh}}$.  
The prior on the degree of the shapelet, $\text{deg}$, follows a discrete uniform distribution $\mathcal{DU}[0,5]$. 
For the shapelet amplitudes, $A_{\text{sh}}$, we employ a conditional uniform prior $\mathcal{CU}[0,A_{\text{sh}}^{\text max}(f_{\text{sh}})]$, where the prior maximal value is determined as a function $f_{\text{sh}}$; the description of this function is found in Appendix~\ref{appx:sh_prior}. This approach is designed to prevent shapelets from fitting spectral lines excessively, thereby restricting their amplitudes to a range suitable for capturing intermediate features in the data.

The priors on $\beta$ and $f_{\text{sh}}$ are different for the broadband (BBS) and the narrowband (NBS) noise components introduced in Eq.~\ref{eq:main_fit}. The parameter $\beta$ follows a log-uniform distribution with different boundaries for each, $\beta^{\text{BBS}} \sim \log \mathcal{U}[1,500]$, and $\beta^{\text{NBS}} \sim \log \mathcal{U}[0.5,10]$. The prior on $f_\text{sh}^{\text{BBS}}$ is a uniform distribution $\mathcal{U}[f_\text{min}, f_\text{max}]$ between the minimum and maximum of the frequency range, while the $f_{\text{sh}}^{\text{NBS}}$ prior follows the same distribution as $f_l$; i.e., an interpolation prior $\texttt{Interp}[f_\text{min}, f_\text{max}]$ described in Appendix~\ref{appx:lines_prior}.   

Lastly, we set $N_{\text{BBS/NBS}}$ to a constant predetermined number. This is because $\text{deg}$ is a free parameter, and if both $N_{\text{BBS/NBS}}$ and $\text{deg}$ were free, it would require two levels of transdimensional sampling, which is not currently supported by the transdimensional extension of \bilby{}. However, when $\text{deg}$ is set to zero, the entire sum of shapelets is ignored, meaning that $\text{deg}$ effectively determines $N_{\text{BBS/NBS}}$.

\begin{table}
    \centering
    \begin{tabular}{lcc}
    \hline
        Parameter & Prior Type & Range \\
        \hline
        $N_\text{PL}$ [-] & $\mathcal{DU}$ & [0,5]  \\
        $A_{\text{PL}}$ [$\mathrm{Hz}^{-\alpha -1/2}$] & $\log \mathcal{U}$ & [$10^{-30}$,$10^{-13}$]   \\
        $\alpha$ [-] & $\mathcal{CU}$  & [-10,2] \\
        \hline
         $N_\text{line}$ [-] & $\mathcal{DU}$ & [0,20] \\
         $\Gamma$ [Hz] & $\log \mathcal{U}$ & $[10^{-3},1]$  \\
         $\zeta$ [-] & $\mathcal{G}_t$ & [0.1,5] \\
        $\tau$ [$\mathrm{Hz}^{-1}$]& $\delta$ & 5.2 \\
        $f_{l}$ [Hz]&  $\texttt{Interp}$ & $[f_\text{min}, f_\text{max}]$ \\
         $A_{l}$ [$\mathrm{Hz}^{-1/2}$]& $\log \mathcal{CU}$ & [-,-]  \\
        \hline
        $N_{\text{BBS/NBS}}$ [-] & $\delta$ & 4 \\
        $\text{deg}_{\text{BBS/NBS}}$ [-] & $\mathcal{DU}$ & [0,5]  \\
        $A_\text{sh}^{\text{BBS/NBS}}$ [-]& $\mathcal{CU}$ & [0,-]  \\
        $\beta_{\text{BBS}}$ [Hz] & $\log \mathcal{U}$ & [1,500] \\
        $\beta_{\text{NBS}}$ [Hz] & $\log \mathcal{U}$ & [0.5,10] \\
        $f_\text{sh}^{\text{BBS}}$ [Hz] &  $\mathcal{U}$ & $[f_\text{min}, f_\text{max}]$  \\
        $f_\text{sh}^{\text{NBS}}$ [Hz] &  $\texttt{Interp}$ & $[f_\text{min}, f_\text{max}]$  \\         
         \hline
    \end{tabular}
    \caption{Table of Priors. Here, $\mathcal{DU}$ indicates a discrete uniform prior, $\mathcal{CU}$ signifies a conditional uniform prior, $\log \mathcal{U}$ represents a log uniform prior, $\mathcal{G}_t$ denotes a truncated Gaussian prior, $\delta$ denotes the Dirac delta function, and $\texttt{Interp}$ stands for an interpolation prior (see Appendix~\ref{appx:preprocessing}). Additionally, $\log \mathcal{CU}$ represents a conditional log uniform prior, and $\mathcal{U}$ denotes a uniform prior. Here, BBS and NBS refer to the broadband and narrowband shapelets, respectively, and should be treated as distinct sets of parameters. Refer to the text for additional details on the priors.}
    \label{tab:prior}
\end{table}

\subsection{Sampling}\label{subsec:sampling}
To manage computational costs, we divide our  procedure into several steps:
\begin{enumerate}
    \item We identify likely lines using the algorithm described in Appendix~\ref{appx:lines_prior} to explore adjacent data.
    \item We perform Bayesian inference only for the broadband noise parameters ($\sigma_{\text{PL}}$ and $\sigma_{\text{BBS}}$), where all the estimated line frequency bins are notched out based on the results of the previous step. 
    \label{item:BB_fit}
    \item The observing band is dynamically divided into typically six frequency sub-bands, following the algorithm described in Appendix~\ref{appx:segmentation}.
    \item For each sub-band, we perform Bayesian inference, while fixing the broadband noise parameters to the maximum-likelihood values of the preferred model obtained in step~\ref{item:BB_fit}. That is, we fit only the narrowband features ($\sigma_{\text{line}}$ and $\sigma_{\text{NBS}}$) within the frequency bins that were notched out in step~\ref{item:BB_fit}.  \label{item:NB_fit}   
    \item We combine the posterior samples from all sub-bands with the broadband posterior samples.
\end{enumerate}
This process is possible because the frequency bins notched out in step~\ref{item:BB_fit} and reintroduced in step~\ref{item:NB_fit} can be treated as independent datasets. Additionally, the $\max_f$ operation introduced in the model construction in Eq.~\ref{eq:main_fit} decouples the broadband and narrowband components, making the model less sensitive to the broadband component in regions dominated by the narrowband component. 
This approach allows each sub-band to use a relatively simple model with not-too-many parameters, while the final output reflects a more complex and comprehensive model. Moreover, performing the sub-band fitting in parallel significantly reduces the total sampling time.
The sampling procedure throughout these steps is carried out using the \dynesty{} sampler~\cite{dynesty}.

The \tpowerbilby{} software package offers flexibility by allowing users to select different levels of inference, enabling rapid estimation of $\sigma(f)$, producing a hybrid solution that combines \tpowerbilby{} and Welch’s method. For additional details, see Appendix~\ref{appx:tPowerbilby}.

\subsection{High-quality data versus low-quality data}\label{sec:quality}

While the noise curve contains lines with very high values of $\sigma(f)$, the sensitivity to astrophysical sources is primarily defined by the broadband component. These order-of-magnitude differences create two distinct classes of frequency bins. We consider frequency bins well described by the broadband as `high-quality data', while those dominated by lines we classify as `low-quality data'. 
Data is considered high-quality if it satisfies 
\begin{align}
|d(f)| < 5 \times  \sum_{i}^{N_{\text{PL}}}\sigma_{\text{PL}}(f_l,\Lambda^{PL}_i) + 
\sum_{l}^{N_{\text{BBS}}}\sigma_{\text{BBS}}(f,\Lambda^{\text{BBS}}_{l})  .
\end{align}
We use this classification of frequency bin quality in later parts of the analysis, where the discarded low-quality data can be treated as notches.
The factor of 5 is selected to strike a balance between rejecting overly noisy frequencies while retaining most of the signal. 
When low-quality data is excluded, the resulting optimal SNR is reduced by $\lesssim 0.1\%$.

\section{Application to LIGO--Virgo data}\label{sec:study_of_real_data}
In this section, we apply our method to three important gravitational-wave events: \gwFirst, the first gravitational-wave signal ever detected~\cite{GW150914}; \gwMay, the most massive black hole binary identified to date~\cite{PhysRevLett.125.101102}; and \texttt{GW190929\_012149} (referred to here as \gwSep), which may contain traces of eccentricity~\cite{GWTC-2.1,GW190929}. 
We show how our method produces an improved description of the noise compared to the commonly used empirical techniques, specifically Welch's method and \texttt{BayesLine}, and we illustrate how changes to the noise model lead to subtle but non-negligible differences in the posterior estimates for astrophysical parameters.

\subsection{Results}\label{subsec:asd_result}
\begin{figure*}
    \centering
    \includegraphics[width=2\columnwidth]{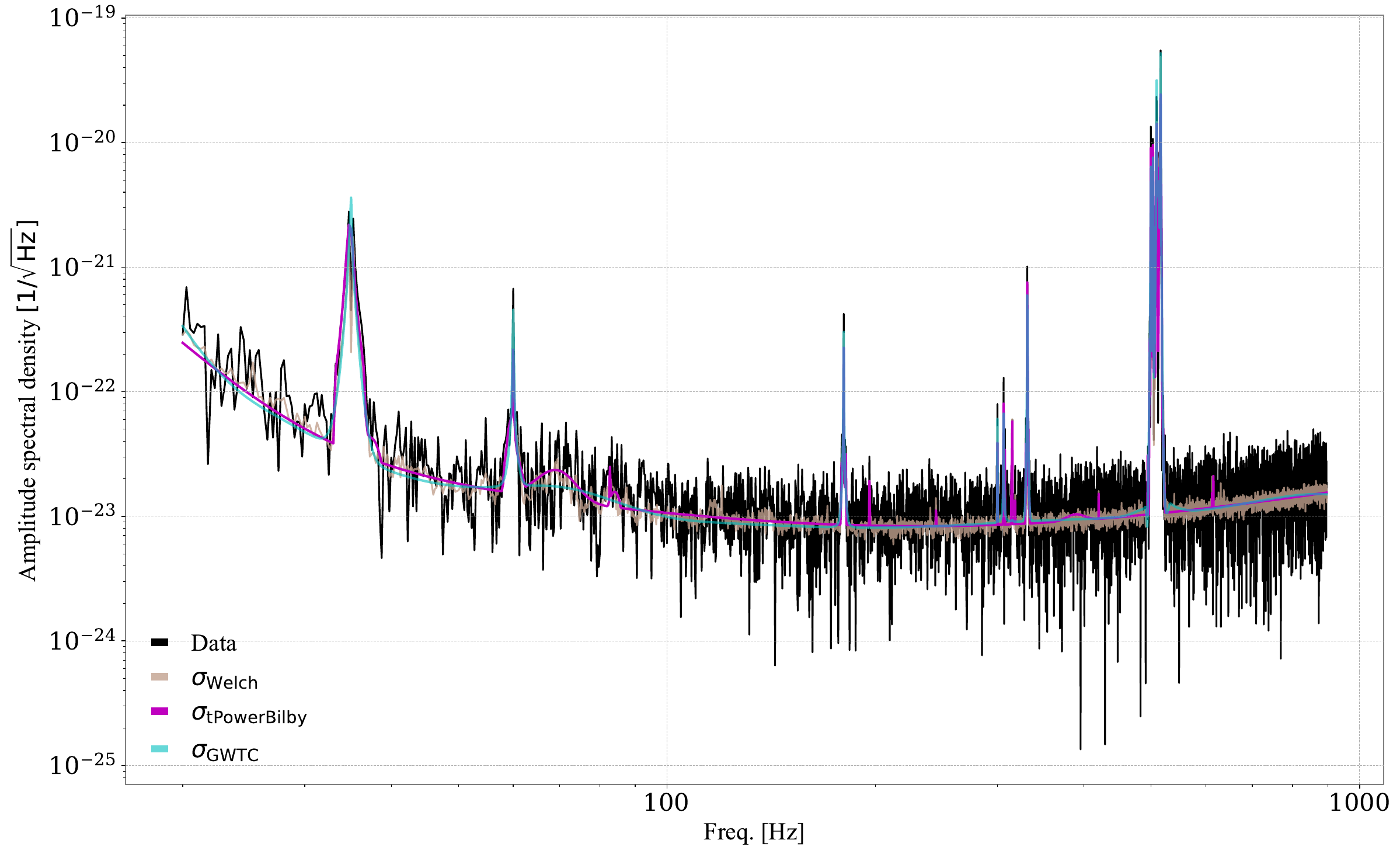}
    \caption{
    Estimates for the noise for the Livingston interferometer in the vicinity of the gravitational-wave signal \gwFirst{}.
    The horizontal axis is frequency while the vertical axis is amplitude spectral density, $\sigma(f)$.
    In black we plot $\unit[4]{s}$ of data taken immediately following the gravitational wave event.
    The beige curve represents Welch's method estimation ($\sigma_{\text{Welch}}$) based on $32 \times 4$ seconds of data. The cyan curve shows the noise estimate used in GWTC-3 ($\sigma_{\text{GWTC}}$). The magenta curve displays the \tpowerbilby{} noise estimate ($\sigma_{\tpowerbilby{}}$) obtained from the maximum-likelihood posterior sample of the preferred model.} 
    \label{fig:gw_150914_L1_asd}
\end{figure*}

\begin{figure*}
    \centering
    \includegraphics[width=2\columnwidth]{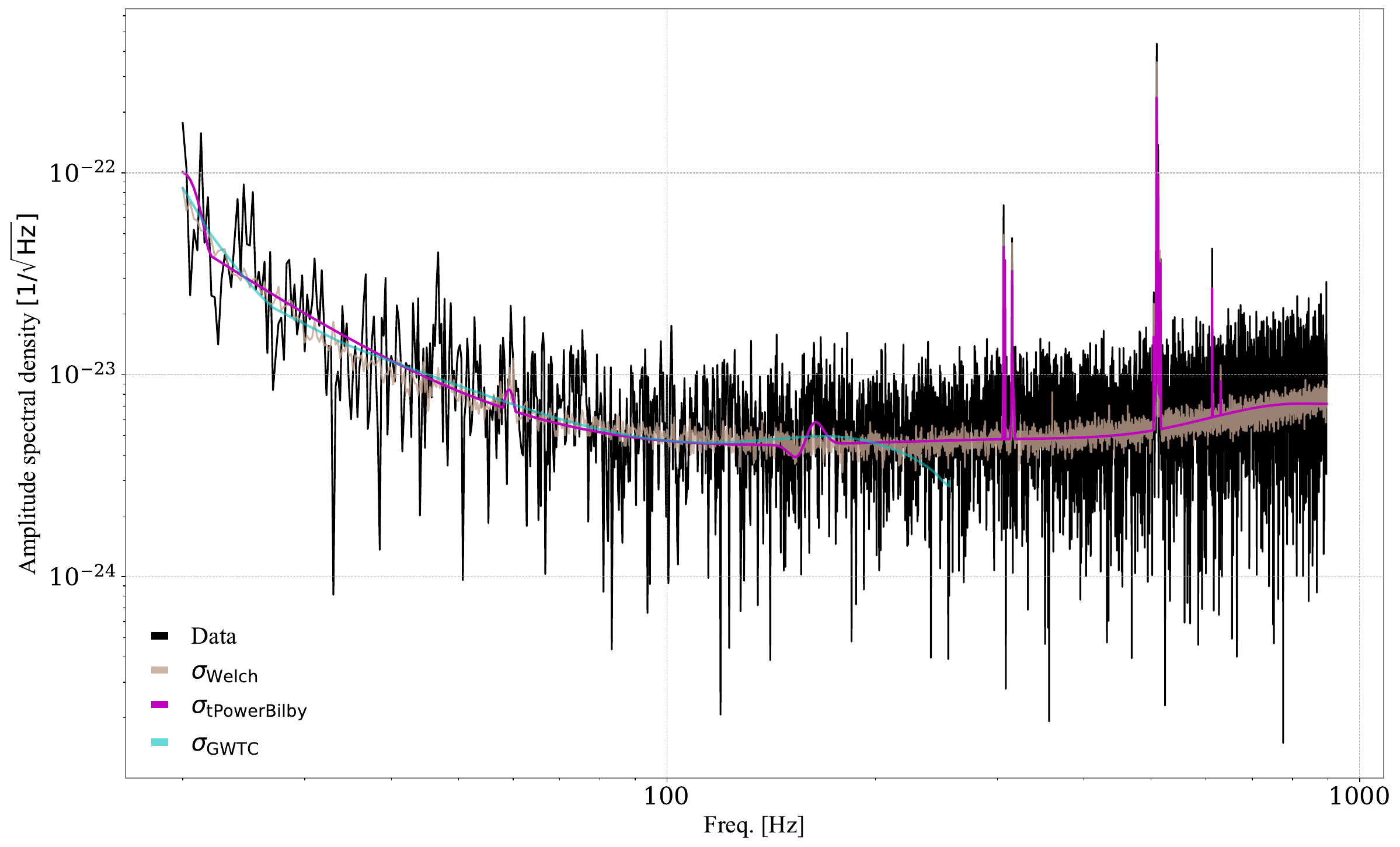}
    \caption{
    Estimates for the noise for the Livingston interferometer in the vicinity of the gravitational-wave signal \gwMay{}. All curves have the same description as Fig.~\ref{fig:gw_150914_L1_asd}. Note that the $\sigma_{\text{GWTC}}$ curve is available only up to 225 Hz. }
    \label{fig:gw_190521_L1_asd}
\end{figure*}

In Figs.~\ref{fig:gw_150914_L1_asd} and~\ref{fig:gw_190521_L1_asd}, we present our $\sigma(f)$ fit for the Livingston interferometer data at times immediately following \gwFirst{} and \gwMay{}, respectively.
The data $|d(f)|$ is shown in black, the beige curve represents the Welch method estimation $\sigma_{\text{Welch}}$, the cyan curve is the noise estimate used in GWTC-3~\citep{GWTC-3}, and the magenta curve represents the maximum-likelihood noise estimation of the preferred model using \tpowerbilby{}. 
The $\sigma_{\text{GWTC}}$ fit was obtained using \texttt{BayesLine} with on-source data and estimated from the median value for each frequency bin's inferred posterior distribution~\cite{GWTC-2.1}.
Qualitatively, all curves are broadly consistent with each other.     

\begin{figure*}
    \centering
    \includegraphics[width=2.0\columnwidth]{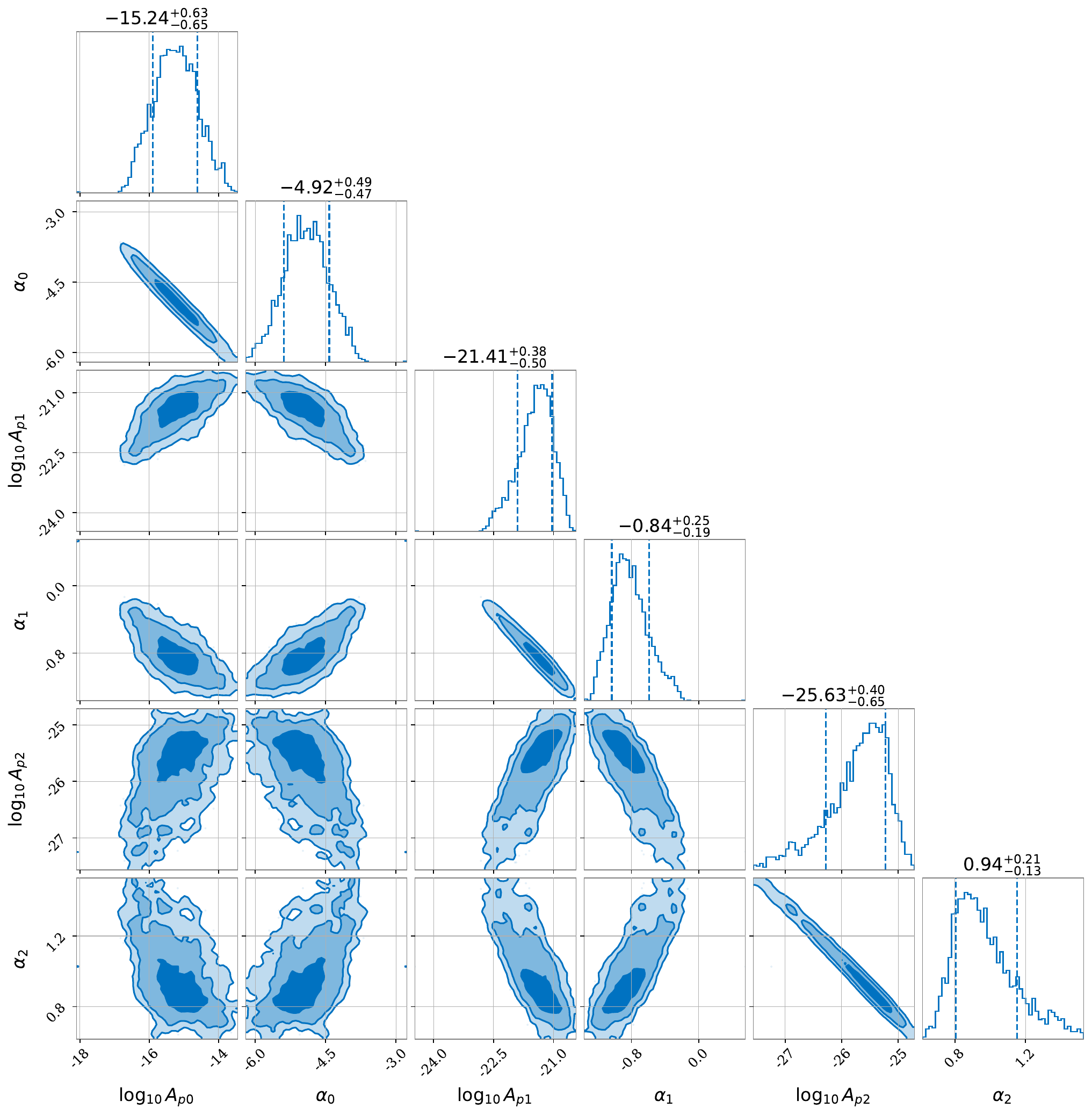}
    \caption{
    Marginalized posterior distribution for the broadband noise parameters for the Livingston interferometer following the \gwFirst{} event.
    The parameters $\alpha_\textit{i}$ are dimensionless, while  $A_{p\textit{i}}$ are in units of $\text{Hz}^{-\alpha_\textit{i}-1/2}$, where $\textit{i}=0,1,2$ denotes the running index of the sum of $\sigma_{\text{PL}}(f,\theta_i^{\text{PL}})$ in Eq.~\ref{eq:sigma_BB_NB}.  
    }  
    \label{fig:gw_150914_L1_power_law}
\end{figure*}

In Fig.~\ref{fig:gw_150914_L1_power_law}, we show the marginalized posterior distribution for the $\alpha$ parameters for the Livingston interferometer for \gwMay{}, where 
the preferred model consists of three power laws. 
The first two spectral indices describe a complex combination of seismic noise, control noise, thermal noise, and squeezer effects, while the third is entirely due to the Fabry-Perot cavity dynamics. One could consider fixing a parameter like $\alpha_3$ to its expected value. If this assumption is reliable—because there always is a power-law component that is well characterized by $\alpha_3 = 1$---then this assumption could potentially improve our fits. The risk of this approach, however, is that a value of $\alpha_3 \neq 1$ might produce a better fit in practice due to imperfections in our noise model. We leave this as a topic for future study.\footnote{In order to fix a parameter in the noise model or set a specific prior boundary, the user may adjust the \tpowerbilby{} configuration file.}

\begin{figure}
    \centering
    \includegraphics[width=\columnwidth]{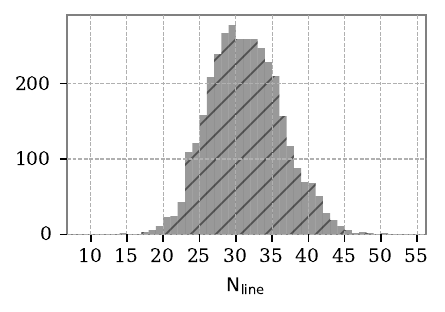}
    \caption{
    Distribution of $N_\text{line}$ for the Livingston interferometer in the vicinity of the \gwMay{} event. This distribution is obtained by randomly combining the various sub-band posterior samples, as described in Sec.~\ref{subsec:sampling}. 
    }  
    \label{fig:gw_190519_n_lines}
\end{figure}

We show the marginalized distribution of the number of lines around \gwMay{} for the Livingston interferometer in Fig.~\ref{fig:gw_190519_n_lines}.
Interestingly, the relatively large width of the distribution suggests that no single model is strongly preferred, highlighting transdimensional sampling as the appropriate tool for this analysis.   

Different observatories have different numbers of lines, and one observatory can have different numbers of lines at different times, typically between 16 and 83.
As a rule of thumb, Hanford and Livingston have fewer lines ($\approx 36$), while Virgo tends to have more lines ($\approx 74$). 
The typical number of shapelets, including both broadband and narrowband shapelets for Hanford and Livingston is $N_\text{sh}\approx 19$, while for Virgo $N_\text{sh}\approx 25$. The number of power law functions strongly favours $N_\text{PL}=2$ and $N_\text{PL}=3$ for all observatories, with a slight preference for the latter. 

\begin{figure}
    \centering
    \includegraphics[width=\columnwidth, height=6cm]{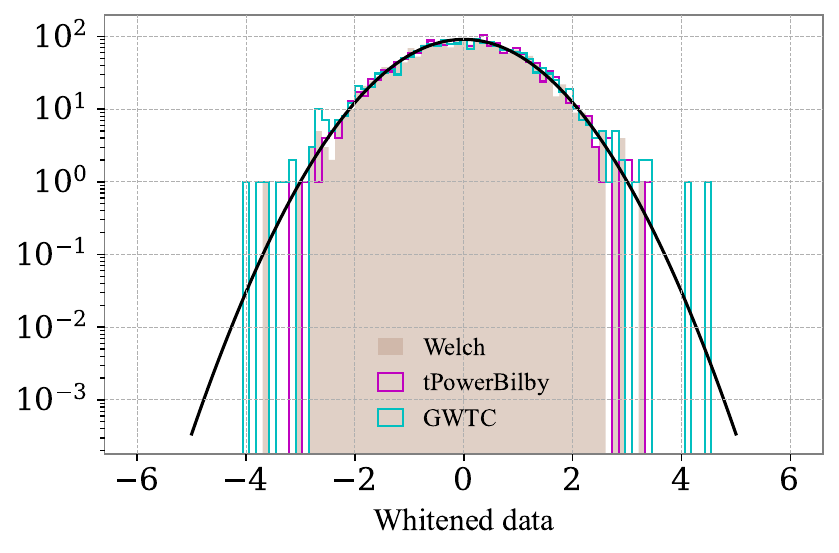}
    \caption{Frequency-domain whitened data preceding the \gwMay{} event from the Livingston interferometer. The beige histogram represents whitened data obtained using Welch’s method. The magenta and cyan curves represent the whitened data obtained using \tpowerbilby{} and the GWTC-3 noise estimate, respectively. The black curve corresponds to the normalized Gaussian distribution.
    }  
    \label{fig:gw_190521_normal}
\end{figure}

In Fig.~\ref{fig:gw_190521_normal}, we show the distribution of the frequency-domain whitened data for Livingston interferometer adjacent to \gwMay{}.
The beige-shaded histogram represents the whitened data using Welch's method, while the magenta and cyan histograms illustrate the results obtained using \tpowerbilby{} and the GWTC-3 noise estimate, respectively. The black curve corresponds to a standard normal distribution. The Kolmogorov-Smirnov $p$-values comparing the distributions to the normalized Gaussian are 0.55 for \tpowerbilby{}, 0.33 for Welch’s method, and 0.23 for GWTC-3 noise estimate. All three $p$-values indicate consistency with a normalized Gaussian distribution, as expected.

\begin{figure*}
    \centering
    \includegraphics[width=2\columnwidth]{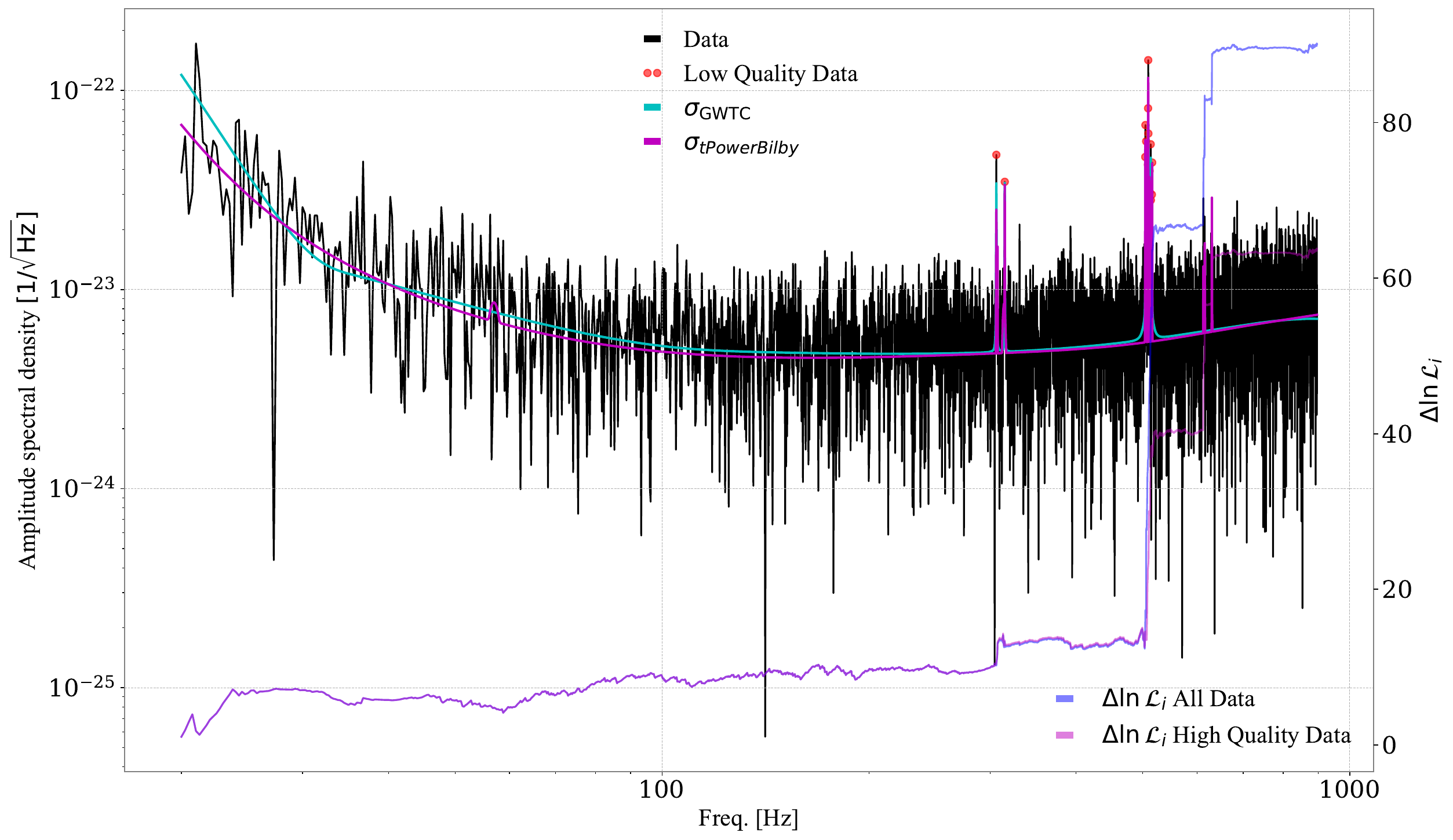}
    \caption{
    Quality of fit spectrum.
  The plot contains two vertical axes with a common horizontal axis. On the left vertical axis, the black curve represents the data prior to the \gwSep{} event for the Livingston interferometer. The magenta curve represents the preferred model maximum-likelihood solution from \tpowerbilby{}, and cyan is the noise curve provided in GWTC-3. The orange markers represent low-quality data (with large values of $\sigma(f)$) that are not included in our goodness-of-fit metrics. On the right vertical axis,  we show the differences in cumulative log likelihood $\Delta \log \mathcal{L}_{i}$ comparing the \tpowerbilby{} fit to the GWTC-3 fit.
  We show the results for two cases: the blue curve is calculated with the entire dataset (including low-quality data), while the red curve is obtained using only high-quality data.
  }
    \label{fig:cumsum_log}
\end{figure*}

Next, we compare our fit to the fit provided in GWTC-3. 
We measure the difference in natural log likelihood to compare our preferred model maximum-likelihood fit to the fit from GWTC-3:
\begin{align}
    \Delta \ln \mathcal{L} = \ln\mathcal{L}^{\tpowerbilby{}} - \ln\mathcal{L}^{\text{GWTC}} .
\end{align}
In Fig.~\ref{fig:cumsum_log}, we plot the cumulative $\Delta \ln {\cal L}$ as a function of frequency for \gwSep{}, where $\Delta\ln\mathcal{L}$ is reported on the right-hand axis.
In blue we show the results using all data: both the high-quality and low-quality data discussed in \ref{sec:quality}.
In red we show the results obtained using only the high-quality data.
We find that \( \Delta \ln \mathcal{L} \) tends to increase across the frequency spectrum, with notable gains associated with spectral lines. In other words, our maximum-likelihood fit is a quantitatively better description of the data than that provided in GWTC-3 and used for astrophysical inferences of those catalog events.
This trend is repeated across different interferometers for the events studied in this manuscript.

Lastly, we report the procedure’s wall time. 
The broadband estimation (the second step in Sec.~\ref{subsec:sampling}) typically takes around 15-30 minutes, whereas the entire sampling procedure can take between 1 to 4 hours. These times are measured using multiprocessing on a high-performance computing cluster.

The sampling time can be reduced by avoiding sampling low-quality data ( i.e., very high values of $\sigma(f)$), which is expected to have a negligible effect on the astrophysical inference. Low quality data create a narrow likelihood surface, posing a challenge for the sampler. This option is available to the user as part of \tpowerbilby{}, as mentioned in Sec.~\ref{subsec:sampling} and Appendix~\ref{appx:tPowerbilby}. Another possible approach to reducing the sampling time is to utilize the JAX package~\citep{jax2018github}, which leverages techniques such as just-in-time compilation and GPU acceleration to optimize computations, which is left for future investigation.

\subsection{Marginalising Over Uncertainty in  \texorpdfstring{\(\sigma(f)\)}{sigma(f)}}\label{subsec:procedure}
In this subsection, we discuss how to apply our estimates of $\sigma(f)$ to astrophysical inference.
The first question we confront is: how much data should we use use to estimate $\sigma(f)$?
In principle, our $\sigma(f)$ fits can be carried out with any length of data.
However, there is a trade off: using relatively short data segments may produce less accurate estimates of $\sigma(f)$ if the noise is stationary, but fits with relatively longer data segments may be less robust due to non-stationary noise.

Experimenting with different variations, we find a noise model that is both robust to non-stationarity and accurate.
We produce \textit{two} fits of $\sigma(f)$---one immediately before the signal, and one immediately after.
Then we model the noise in the segment containing the signal as a frequency-dependent mixture model of these two noise estimates. 
The likelihood function is:
\begin{align}\label{eq:mixture}
    {\cal L}(d | \theta) = 
    \int d\Lambda^\text{pre} \int d\Lambda^\text{post}
    \prod_k & 
    \big[\lambda \, \mathcal{L}(d_k | \theta,\Lambda^\text{pre}) 
    + \nonumber\\ 
    &  (1-\lambda)\mathcal{L}(d_k | \theta,\Lambda^\text{post})\big].
\end{align}
Here, $d$ is the data and $\theta$ denotes the gravitational-wave signal parameters. The variables $\Lambda^\text{pre}$ and \( \Lambda^\text{post}\) are the \tpowerbilby{} estimates for the noise parameters before and after the gravitational wave event, respectively. 
The product is over frequency bins $k$.
The variable $\lambda$ is a mixing hyperparameter, which can be fit as a free parameter or which can be tuned as part of the model. 
In this work, we adopt a fiducial value of $\lambda=0.5$.
The variable $\mu$, which appears explicitly in the likelihood definition in Eq.\ref{eq:likelihood}, is the gravitational waveform; we use the \texttt{IMRPhenomXPHM} approximant~\cite{approximant}. 

The mixture model is designed to account for non-stationary noise observed in the 4-second data segment. Testing the robustness of the estimated noise curves on adjacent time stretches reveals sharp drops in \( \Delta \ln \mathcal{L} \) relative to Welch's method, which we source back to non-stationary noise. By incorporating the mixture model, we mitigate most of these drops (see Appendix~\ref{appx:log_l} for an example). 

To highlight the usefulness of the mixture model, we calculate
\begin{align}
    \Delta \ln \mathcal{L} =  \ln\mathcal{L}^{\tpowerbilby{}} - \ln\mathcal{L}^{\text{Welch}} ,
\end{align}
comparing the maximum likelihood estimates from \tpowerbilby{} with Welch's method over a $\unit[4]{s}$ time window following the gravitational wave event and using high-quality data.
The results are summarized in Tab.~\ref{tab:log_l}.  
The mixture model yields $\Delta\ln {\cal L}$ values ranging from -21 to 49, consistently outperforming the pre- and post-event estimates.

\begin{table}[h!]
\centering
\begin{tabular}{|l|c|c|c|c|c|c|c|c|c|}
\hline
interferometer & \multicolumn{3}{c|}{Livingston} & \multicolumn{3}{c|}{Hanford} & \multicolumn{3}{c|}{Virgo} \\
\cline{1-10}
 Event & Pre & Post & Mix & Pre & Post & Mix  & Pre & Post & Mix \\
\hline
\gwFirst{} & 8.9 & 34 & 37 & -37 & -67 & -21 & - & - & - \\
\gwMay{} & 8.3 & 15 & 32 & 17 & 12 & 44 & -11 & 0.77 & 31 \\
\gwSep{} & 40 & 37 & 49 & -34 & 18 & 36 & -51 & -41 & -7.5 \\
\hline
\end{tabular}
\caption{$\Delta \ln \mathcal{L}$ estimates comparing \tpowerbilby{} and Welch’s method on high-quality data following each gravitational wave event. The estimation is conducted three times using \tpowerbilby{}: pre-event, post-event, and a mixture model of the two.
Data from the Virgo observatory are not available for \gwFirst.
}
\label{tab:log_l}
\end{table}

Having defined our noise model, we use importance sampling to reweight posterior samples for the astrophysical parameters $\theta$ obtained using a $\sigma_{\text{GWTC}}(f)$ obtained with \texttt{BayesLine}~\cite{reweight2}.
This allows us to separate the problem of noise estimation from the problem of astrophysical parameter estimation, which helps reduce the computational cost.
This approach works so long as the ``proposal distribution'' calculated with \texttt{BayesLine} is sufficiently similar to the ``target distribution'' obtained with our noise model \citep{reweight2}.
We calculate weights for each posterior sample $\theta_i$ by marginalising over the noise parameters $\Lambda$:
\begin{equation}
w_i = \frac{1}{N} \sum_{k} \frac{\mathcal{L}(d|\theta_i,\Lambda_k)}{\mathcal{L}(d|\theta_i,\sigma_{\text{GWTC}})},
\end{equation}
where $k$ runs over the posterior samples for the noise model.

In order to check if the proposal distribution is sufficiently similar to the target distribution, we calculate the re-weighting efficiency \citep{reweight2}: 
\begin{equation}
\epsilon =  \frac{1}{n} \frac{\left(\sum_{i}^{n} w_i\right)^2}{\sum_{i}^{n} w_i^2}.
\end{equation}
An efficiency $\gtrsim 2\%$ is typically sufficient to produce well-converged posteriors for the target distribution.
The natural log Bayes factor is simply the average weight:
\begin{equation}
\ln \mathcal{B} = \ln \overline{w} = \ln \frac{\mathcal{Z}_{\Lambda}}{\mathcal{Z}_\text{GWTC}}.
\end{equation}

We calculate $\ln {\cal B}$ two ways: once using all of the data and once using only the high-quality data described in~\ref{sec:quality}.
Since the low-quality data is associated with large $\sigma(f)$, it does not contribute significantly to the inference of astrophysical parameters, though, it can affect the evidence value.
By calculating the $\ln {\cal B}$ for high-quality data, we can ascertain the quality of fit on the data most relevant for astrophysics.

In Appendix~\ref{appx:injection_study}, we present an injection study following the procedure described here and in Sec.~\ref{subsec:procedure}, validating \tpowerbilby{} noise estimate.  

\subsection{Astrophysical Inference Results}\label{subsec:PE_results}
\begin{figure*}
    \centering
    \includegraphics[width=2\columnwidth]{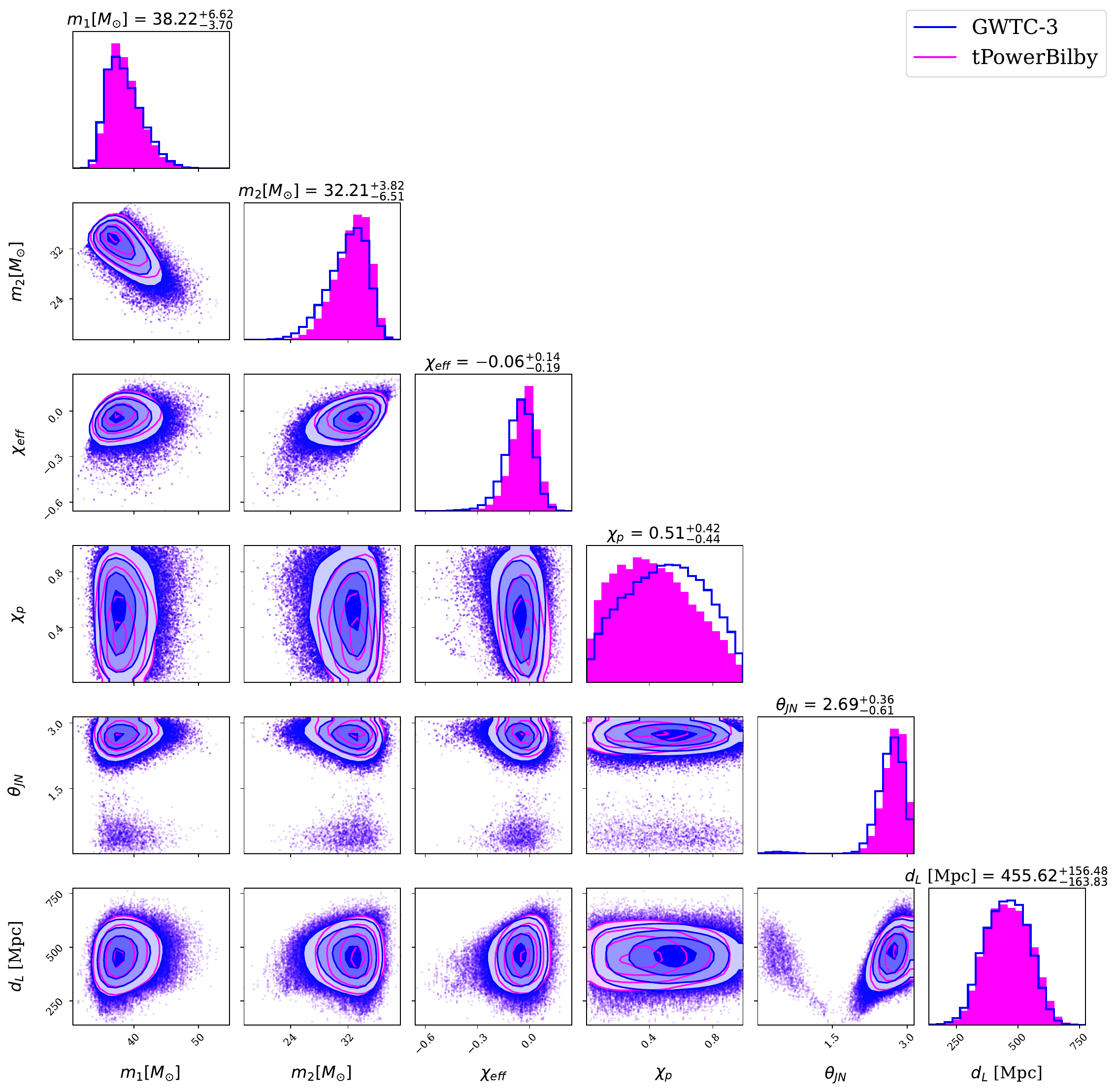}
    \caption{
    Marginalized posterior distributions for \gwFirst{} displaying the masses and spins of the black holes, the inclination angle, and the distance to the binary system. GWTC-3’s posterior samples are shown in purple, while \tpowerbilby{} weighted posterior samples are represented in magenta.
    }  
    \label{fig:gw_150914_corner}
\end{figure*}

\begin{figure*}
    \centering
    \includegraphics[width=2\columnwidth]{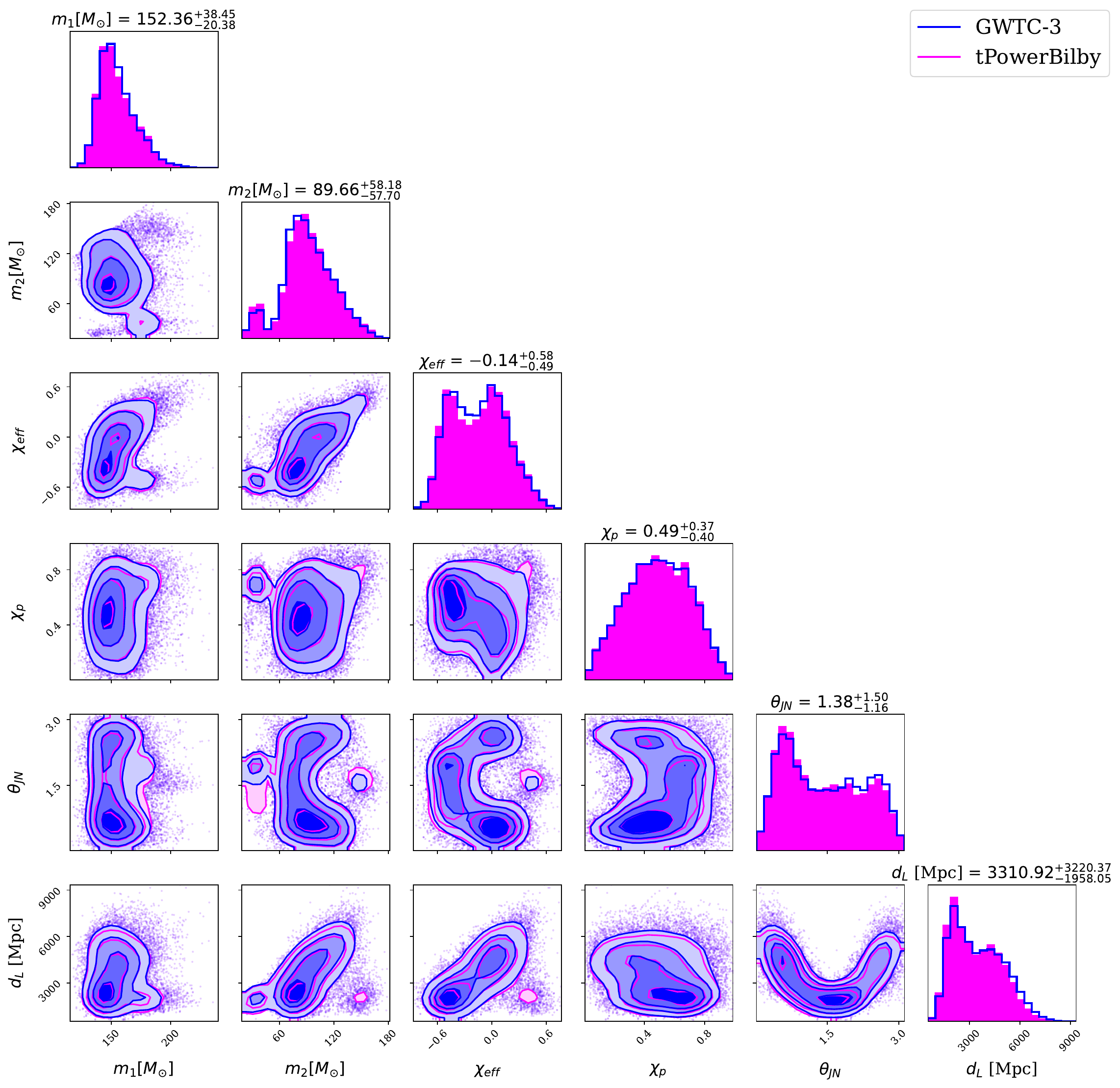}
    \caption{
      As with Fig.~\ref{fig:gw_150914_corner}, but for the \gwMay{} event.}  
    \label{fig:gw_190521_corner}
\end{figure*}

\begin{figure*}
    \centering
    \includegraphics[width=2.0\columnwidth]{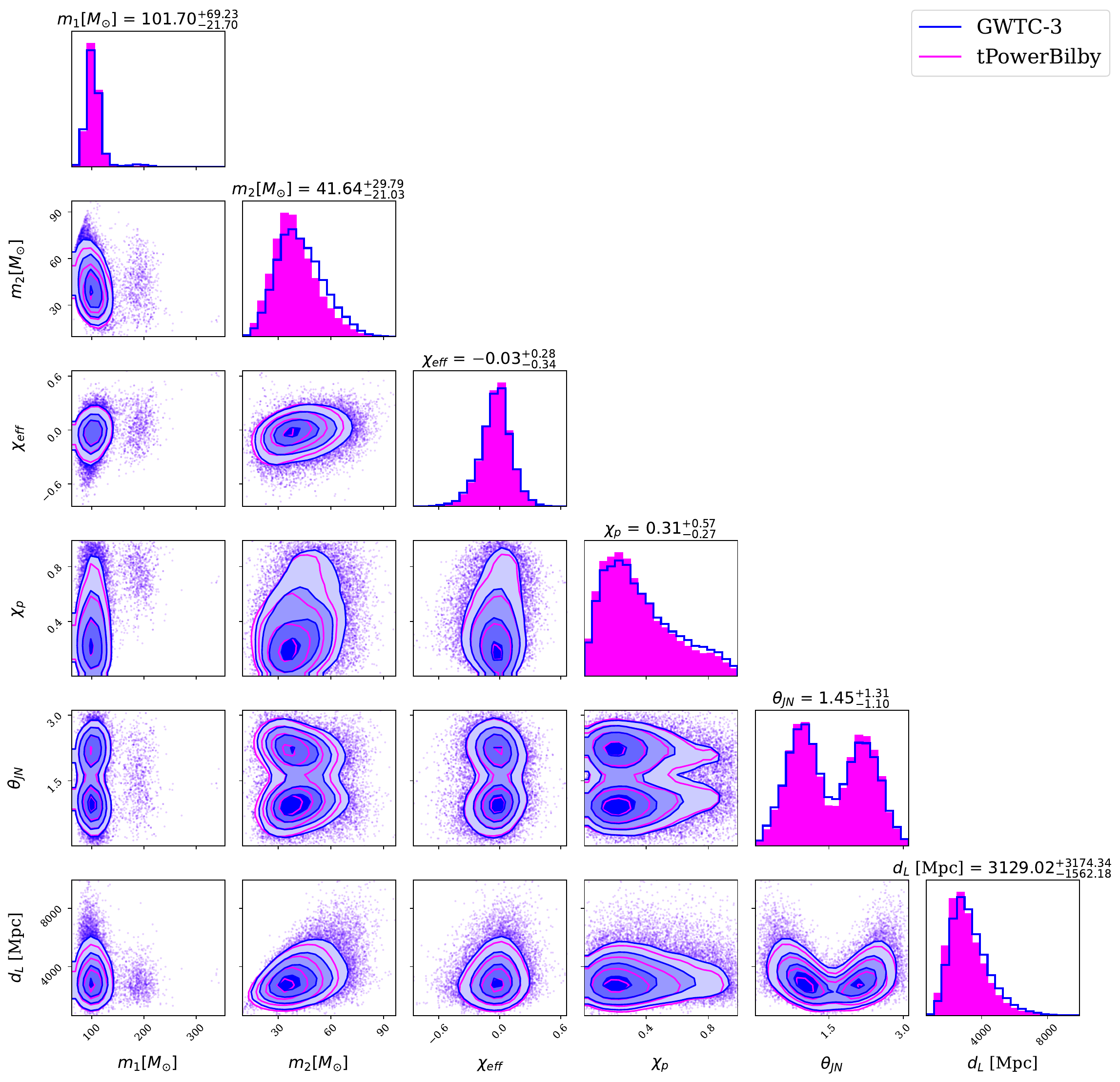}
    \caption{
    As with Fig.~\ref{fig:gw_150914_corner}, but for the \gwSep{} event.}  
    \label{fig:gw_190929_corner}
\end{figure*}

In Figs~\ref{fig:gw_150914_corner},~\ref{fig:gw_190521_corner}, and~\ref{fig:gw_190929_corner}, we show marginalized posterior distributions for the astrophysical parameters of three events: \gwFirst{}, \gwMay{}, and \gwSep{}.
In each case we use the approximant \texttt{IMRPhenomXPHM}.
The results obtained from GWTC-3 are shown in purple while the results obtained with our noise model are shown in magenta. 
We carry out inference for all astrophysical parameters, but we only show here marginalized distributions for the masses, spins, inclination angles and luminosity distances.

The most significant effect is observed for \gwFirst{}, where the changes in the marginal distribution of the effective precession spin
parameter $\chi_p$ are evident in the corner plots. These include $\approx 7\%$ change in the $90\%$ credible interval and $\approx 10\%$ shift in the median value, driven by lower values seen in both component spins $\chi_{1}$ and $\chi_{2}$. Minor changes, around $2\%$, are observed for other parameters, with the effective inspiral spin parameter $\chi_{\text{eff}}$ being the most notable.

For \gwMay{}, the changes in the distributions are relatively smaller. The most significant impact is on the luminosity distance $d_{\text{L}}$, which shows $\approx 4\%$ change in the $90\%$ credible interval.
In the case of \gwSep{}, the most significant changes are seen in the distance $d_{\text{L}}$ and the secondary mass $m_2$, both showing $\approx 6\%$ change in the $90\%$ credible interval and $\approx 3\%$ shift in their median values. A smaller change, approximately $3\%$, is observed for $\chi_p$.
  
Table~\ref{tab:results_summary} states the re-weighting efficiency and $\ln \mathcal{B}$ for each event using high-quality data.
The reasonably large efficiency values $39-55\%$ suggests that the target distributions are well converged.
The natural log Bayes factor 
compares the median \texttt{BayesLine} fit, obtained using just the on-source data, to the \tpowerbilby{} results marginalised over noise uncertainty using off-source data as a prior.
The values range from $\ln{\cal B}=8.8$ to 20, implying \tpowerbilby{} provides a significantly better explanation of the data than the GWTC-3 fits.

Comparing the credible intervals in Figs.~\ref{fig:gw_150914_corner}-\ref{fig:gw_190929_corner}, we observe small but non-negligible changes in the credible intervals.
The median values of some parameters change by as much as 10\% while the credible interval widths change by as much as 7\%.
These shifts provide some estimate for the magnitude of systematic error from misspecified noise modelling.
We highlight that they are on the same scale as (or more important than) other major sources of systematic error including: calibration uncertainty~\citep{reweight2}, waveform systematics~\citep{systematics_GW150914}, finite-duration effects \citep{windows}.

\begin{table}[h]
    \centering
    \begin{tabular}{|c|c|c|}
        \hline
        Event & Weighting Efficiency ($\%$) & $\ln \mathcal{B}$ \\ \hline
        \gwFirst{} & 39 & 20 \\ \hline
        \gwMay{} & 45 & 8.8 \\ \hline
        \gwSep{} & 55 & 20\\ \hline
    \end{tabular}
    \caption{Summary of the gravitational wave event analyses, presenting the weighting efficiency and $\ln \mathcal{B}$ values comparing the $\sigma_{\tpowerbilby{}}$ and $\sigma_{\text{GWTC}}$ models on high-quality data.
    }
    \label{tab:results_summary}
\end{table}

\section{Discussion and Conclusion}\label{sec:conclusions}
In this study, we implement a transdimensional model for noise estimation with \bilby. 
The code for our model is available on git hub in the \tpowerbilby{} repository.
We illustrate our framework using three events from GWTC-3.
We show that our noise model provides an improved fit over the commonly used Welch's method.
Applying our noise model to gravitational-wave signals, we find non-negligible shifts in the posterior distributions of astrophysical posteriors suggesting that systematic error from misspecified noise is arguably the largest source of systematic error in astrophysical inference for some events and parameters. However, for the intrinsic parameters of very high-mass events, waveform systematics can be more significant~\citep{GW190521_prop}. 
Since our model only attempts to account for \textit{stationary} noise, we speculate that the true systematic error from misspecified noise modelling is likely much larger than our estimates.
           
These systematic errors could have significant implications for analyses that combine many events, such as population studies where small systematic errors may combine to produce errors large enough to yield faulty astrophysical inferences; see, e.g., \citep{o2_pop,o3a_pop,gwtc-3_pop}. 
Achieving a more accurate measurement of the Hubble constant~\cite{hubble2017} is another example. 
Tests for general relativity are also susceptible to various systematic effects; see ~\cite{gupta2024possiblecausesfalsegeneral}. 

Ref.~\citep{PhysRevD.106.104017} highlights the importance of both robust noise modeling and marginalization over the noise model, as well as its impact on the inferred astrophysical parameters especially for subtle effects such as spin-precession.

Ref.~\cite{Biscoveanu_2020} uses the noise estimation from \texttt{BayesLine} to assess its impact on parameter estimation. The authors draw $\sigma(f)$ estimates from \texttt{BayesLine}'s posterior samples and marginalize over uncertainty in the noise model, performing a similar analysis to the one conducted in this paper. 
Their study involved running approximately 200 parameter estimation samples and found an effect of around $5\%$ in the credible interval width. 

Ref.~\cite{talbot2020gravitationalwave} studies the effect of marginalizing over noise uncertainty, when $\sigma(f)$ is estimated with Welch's method. 
The authors find a strong preference for the marginalized model over the non-marginalized one, with improvements of up to $\ln \mathcal{B} \approx 90$. 
Our results are consistent with both \cite{talbot2020gravitationalwave} and \cite{Biscoveanu_2020}.
All three draw attention to the fact that systematic error from misspecified noise can be a dominant source of systematic error in astrophysical inference.

Ref.~\cite{Maximum_entropy} employs a different methodology for characterizing the noise curve, using maximum entropy spectral analysis on on-source data. This approach finds a consistent estimated uncertainty on the order of $\approx 10\%$ when marginalising over the noise.
In contrast, Ref.\citep{glitch_mitigations2} and Ref.\citep{PSD_effect_on_PE_1} report much smaller values. Specifically, Ref.~\citep{PSD_effect_on_PE_1} estimates an effect on the order of $\approx 2\%$ and concludes that uncertainty in noise estimates is a subdominant factor compared to other sources of uncertainty.

Our noise model, and the implementation in \tpowerbilby{} is heavily influenced by the \texttt{BayesLine} algorithm~\cite{Littenberg_2015}. 
The methods share several similarities: both rely on transdimensional sampling, utilise essentially the same Lorentzian line model, and employ smooth functions to describe the broadband noise. Additionally, both allow for marginalization over uncertainties in the noise model.
However, there are some of the differences.
First, \texttt{BayesLine} employs a customized reverse jump Markov Chain Monte Carlo whereas our algorithm is designed to work with off-the-shelf samplers available to \texttt{Bilby} (for this paper, we used \texttt{Dynesty}).
Second, \texttt{BayesLine} uses splines while our approach uses power laws. 
Third, while \texttt{BayesLine} uses splines to capture complicated spectral features, \tpowerbilby{} uses shapelets. 
Lastly, \texttt{BayesLine}, when combined with \texttt{BayesWave}~\citep{BayesWave}, allows for the estimation of noise from on-source data and marginalization over noise uncertainties whereas we emphasize the use of off-source data.
Ref.~\citep{PSD_effect_on_PE} compares \texttt{BayesLine} with Welch’s method and finds that the data whitened using \texttt{BayesLine} is more consistent with a normalized Gaussian distribution.

Our goals for future work include conducting a systematic extended study of all the events in GWTC-3, followed by a population study to quantify the aggregated effect of the \tpowerbilby{} noise model. Additionally, we aim to further develop the capabilities of \tpowerbilby{} to utilize on-source data and simultaneously sample both the noise and the signal. 
This approach not only provides a more accurate representation during the signal duration but also allows for explicit marginalization over the noise uncertainties. Lastly, a long-term vision is to introduce non-stationary noise handling capabilities to the \tpowerbilby{} framework. This could potentially include glitch mitigation, similar to \texttt{BayesWave}~\citep{BayesWaveUpdated2024,glitch_mitigations0,glitch_mitigations1,glitch_mitigations2}, or addressing other typical non-Gaussian noise sources~\citep{non_gaussian_noise} in order to provide a robust tool for gravitational-wave science.

\section*{acknoweledgements}
This material is based upon work supported by NSF's LIGO Laboratory which is a major facility fully funded by the National Science Foundation.
This work is supported through Australian Research Council (ARC) Centres of Excellence CE170100004, CE230100016, Discovery Projects DP220101610 and DP230103088, and LIEF Project LE210100002.
We thank Lucy Thomas, Katerina Chatziioannou, Meg Millhouse, and Colm Talbot for feedback on an earlier draft of this manuscript. The authors would like to thank the anonymous reviewer for the insightful comments and feedback, which greatly strengthened the manuscript.
The authors are grateful for for computational resources provided by the LIGO Laboratory computing cluster at California Institute of Technology supported by National Science Foundation Grants PHY-0757058 and PHY-0823459, and the Ngarrgu Tindebeek / OzSTAR Australian national facility at Swinburne University of Technology.
LIGO was constructed by the California Institute of Technology and Massachusetts Institute of Technology with funding from the National Science Foundation and operates under cooperative agreement PHY-1764464. This paper carries LIGO Document Number LIGO-P2400574.

\appendix 
\section{\tpowerbilby{} Package}\label{appx:tPowerbilby}

The \tpowerbilby{} package is a user-friendly tool designed to estimate the amplitude spectral density $\sigma(f)$ of ground-based gravitational-wave observatories. 

Based on the user input provided, which includes configuration parameters such as the GPS trigger time and the name of the interferometer, the software automatically retrieves the required data and executes the procedures outlined in this paper.

The software comprises three main components: a preprocessing stage that constructs the priors for the subsequent steps, a sampling stage, and a post-processing stage that handles the results from the various stages and merges them to produce the final estimation.

To maximize flexibility regarding computational costs, the software outputs results after each sampling stage, allowing users to access the sampling results as they become available. In addition, users can choose to fit the entire dataset, focus on broadband data, or analyze specific segments of high-quality data. When choosing not to fit the entire dataset, the software employs the Welch method to estimate missing values, effectively merging \tpowerbilby{} estimates with it to produce usable $\sigma$ samples. A detailed explanation of the available configurations, along with examples of how to use the software, can be found at: \giturl{}.

\section{Preprocessing Algorithms}\label{appx:preprocessing}
The preprocessing stage involves a data-driven construction of the lines and shapelet priors, as well as the division of the entire frequency band into smaller sub-band regions and the identification of high-quality frequency bins. The following subsection outlines the preprocessing algorithms.

\subsection{Line Priors Related Construction}\label{appx:lines_prior}

We begin by estimating the noise using Welch's method, denoted as $\sigma_{\text{Welch}}(f)$, for the data preceding the analyzed segment. Next, we calculate a moving median for $\sigma_{\text{Welch}}(f)$ over a range of 100 frequency bins, which serves as a rough estimation of the broadband component of $\sigma(f)$, labeled as $\sigma(f)^{\text{BB}}_{\text{med}}$. However, the low-frequency range can often be misleading due to edge effects. To mitigate this issue, we employ a RANSAC algorithm~\cite{RANSAC} to fit a single power law in the frequency range below 30 Hz, which provides an approximation of the broadband component in that frequency range. Subsequently, we construct the line locations prior by including every frequency bin (and its two adjacent bins) with equal probability that satisfies the condition 
\[
\sigma_{\text{Welch}}(f) > 3.5 \times \sigma(f)^{\text{BB}}_{\text{med}}. 
\]
Finally, we normalize the distribution to form a proper probability distribution.
We select 3.5 as it roughly corresponds to 3.5 standard deviations of a normal distribution, which is above the $99.9\%$ threshold, indicating that the observed values are unlikely to be random fluctuations in the data.

Next, we reevaluate the broadband noise using 4 seconds of data preceding the analysed segment. First, we remove the frequency bins included in the line locations prior. Then, we fit the remaining data using a sum of power laws, as outlined in Eq.~\ref{eq:power_laws}. The maximum likelihood obtained from this fit serves as our new estimate of the broadband noise, denoted as $\sigma_{\text{Fit}}^{\text{BB}}$. 

We proceed to construct the maximum and minimum for the line amplitude prior, as mentioned in Sec.~\ref{subsec:priors}. We collect 32 time segments of the same duration as the analysed data, and preceding it, denoted as ($\text{data}_{\text{pre}}(f)$). We define the maximum amplitude as a function of frequency by performing the maximum operation among the 32 values 
\[
A_{l}^{\text{max}}(f) = \max_f(\text{data}_{\text{pre}}(f)),
\]
where $\max_f$ indicates the maximum operation per frequency bin. For the minimum amplitude we take the maximal value between $\sigma_{\text{Fit}}^{\text{BB}}(f)$ and the minimum operations among the 32 values, 
\[
A_l^{\text{min}}(f) = \max_f\left(\sigma_{\text{Fit}}^{\text{BB}}(f), \min_f(\text{data}_{\text{pre}}(f)) \right), 
\]
where $\min_f$ indicates the minimum operation per frequency bin. This ensures that no line has an amplitude smaller than the estimated broadband noise.

\subsection{Shapelets Priors Related Construction}\label{appx:sh_prior}
For the shapelets, we define the maximum amplitude as a function of frequency using the same quantity \(\sigma(f)^{\text{BB}}_{\text{med}}\) as defined in Appendix~\ref{appx:lines_prior}. The maximum amplitude is given by 
\[
A_{\text{sh}}^{\text{max}}(f) = 3.85 \times \sigma(f)^{\text{BB}}_{\text{med}}.
\]
The value 3.85 results from \(1.1 \times 3.5\), which incorporates a \(10\%\) safety margin on top of the \(3.5\) factor introduced in Appendix~\ref{appx:lines_prior}.

\subsection{High-Quality Data in Pre-Processing}
As mentioned in Appendix~\ref{appx:tPowerbilby}, \tpowerbilby{} allows for the selection of three levels of data to fit. The second level focuses exclusively on high-quality data, which reduces computation time while retaining the most critical data points. We define high-quality data as 
\[
\text{data}_{\text{HQ}} = \text{data} < 5 \times \sigma_{\text{Fit}}^{\text{BB}}(f),
\]
while also keeping low-frequency data below \(\unit[40]{Hz}\). Here, \(\sigma_{\text{Fit}}^{\text{BB}}(f)\) is defined in Appendix~\ref{appx:lines_prior}.

\subsection{Segmentation Algorithm}\label{appx:segmentation}

The segmentation process aims to divide $\sigma(f)$ frequency range into multiple regions, each subject to further analysis. It relies on the availability of the lines location prior described in Appendix~\ref{appx:lines_prior}. Typically, this prior is sparse, and the analysis leverages this by calculating the difference between adjacent non-zero probability values, effectively identifying empty sections between lines. Differences deemed too small, determined by user input (defaulting to 30 Hz), are discarded. The midpoint of each empty section is considered a candidate for splitting the spectrum at its frequency value. If two values are too close to each other, as specified by the user (defaulting to 100 Hz), the one within the smaller section is discarded. Naturally, the boundaries of the spectrum are fixed points that must be included in the final list of splitting frequency values. Based on this list, $\sigma(f)$ can be divided into multiple regions for further analysis.

\section{\texorpdfstring{$\Delta\log\mathcal{L}$}{Delta log L} Example}\label{appx:log_l}

\begin{figure*}
    \centering
    \includegraphics[width=2.0\columnwidth]{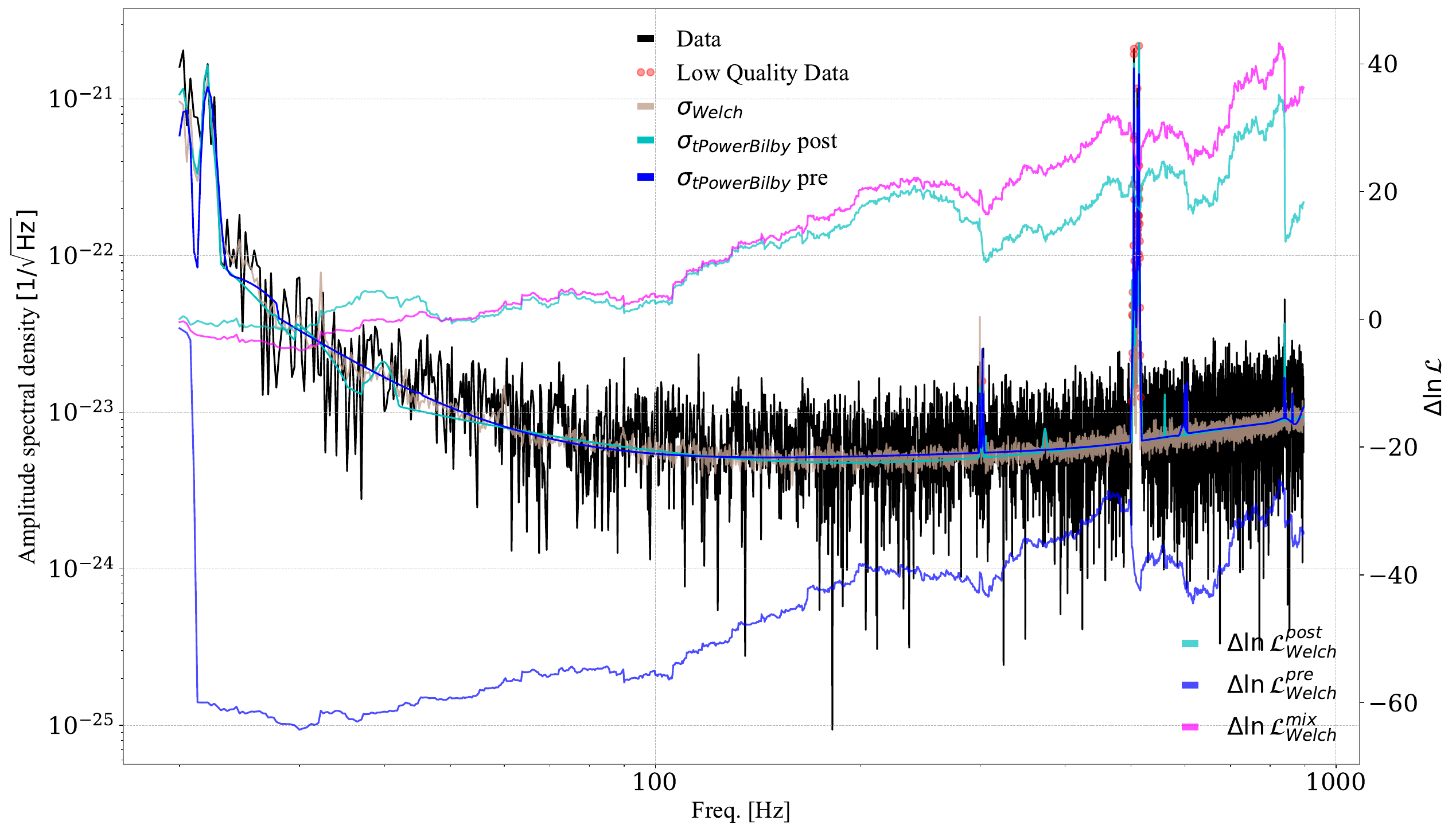}
    \caption{
    Quality of fit spectrum.
  The plot contains two vertical axes with a common horizontal axis. On the left vertical axis, the black curve represents the data preceding the \gwSep{} event for the Hanford interferometer. The cyan curve represents the maximum-likelihood solution of the preferred model from \tpowerbilby{} calculated using data following the \gwSep{} event. The purple curve is based on data preceding the \gwSep{} event. The beige curve is the noise curve estimated using Welch's method. The orange markers represent low-quality data (with large values $\sigma(f)$), which are not included in our goodness of fit metrics. On the right vertical axis,  we plot the differences in cumulative log likelihood $\Delta \ln \mathcal{L}_{i}$ comparing the \tpowerbilby{} fit to the Welch fit.
  We present the results for three cases: the cyan curve is calculated using data following the \gwSep{} event, the purple curve is based on data preceding the \gwSep{} event, while the magenta curve is obtained using the mixture model defined in Eq.~\ref{eq:mixture}.
    }
    \label{fig:log_l_H1}
\end{figure*}

Here we provide an example demonstrating how non-stationarity affects noise estimation and how the mixture model defined in Eq.~\ref{eq:mixture} mitigates this effect.   
For this example, we use data surrounding the \gwSep{} event from the Hanford observatory. In Fig.~\ref{fig:log_l_H1}, we show an estimation of the noise curve and the corresponding calculation of
\[ \Delta \ln \mathcal{L} =  \ln\mathcal{L}^{\tpowerbilby{}} - \ln\mathcal{L}^{\text{Welch}}. \] Specifically, the noise data preceding the \gwSep{} event by 10 seconds is shown in black, while the cyan curve represents the preferred model maximum-likelihood solution from \tpowerbilby{} using data following the event. The blue curve represents the preferred model maximum-likelihood solution from \tpowerbilby{} using data preceding the event. The beige curve indicates the noise curve estimated using Welch’s method. Orange markers represent low-quality data (with large values of $\sigma(f)$), which are not included in the $\Delta \ln \mathcal{L}$ calculation. We present three evaluations of $\Delta \ln \mathcal{L}$: the cyan curve is calculated using data following the \gwSep{} event, the purple curve is based on data preceding the \gwSep{} event, and the magenta curve is obtained using the mixture model defined in Eq.\ref{eq:mixture}.
   
Significant drop in $\Delta\ln\mathcal{L} \approx 50$ is visible in the low-frequency range, below 30 Hz for the $\sigma_{\tpowerbilby{}}$ noise evaluation preceding the event. This drop indicates the presence of non-stationary noise. However, the application of the mixture model results in a significant improvement.
Additional, smaller drops are observed across the entire frequency range; many of these are mitigated, leading to a higher $\ln \mathcal{L}$.    
The figure illustrates a typical scenario where the performance of our model is comparable to that of Welch’s method below 100 Hz. Most of the improvement in $\Delta \ln\mathcal{L}$ occurs at higher frequencies.

\section{Injection study}\label{appx:injection_study}
To validate \tpowerbilby{}, we present an injection study where a binary black hole signal, with  parameters given in Table~\ref{tab:injection_parameters}, is injected into noise strain data recorded 30 seconds after \gwFirst{}.

Using the \tpowerbilby{} noise estimate obtained from a fit to the data immediately following \gwFirst{} (see Sec.\ref{subsec:asd_result} for details), we estimate the astrophysical parameters. Specifically, we utilize the preferred model maximum likelihood curve as the noise model. We then marginalize over the noise uncertainties following the procedure outlined in Sec.\ref{subsec:procedure}.

\begin{figure*}
    \centering
    \includegraphics[width=2\columnwidth]{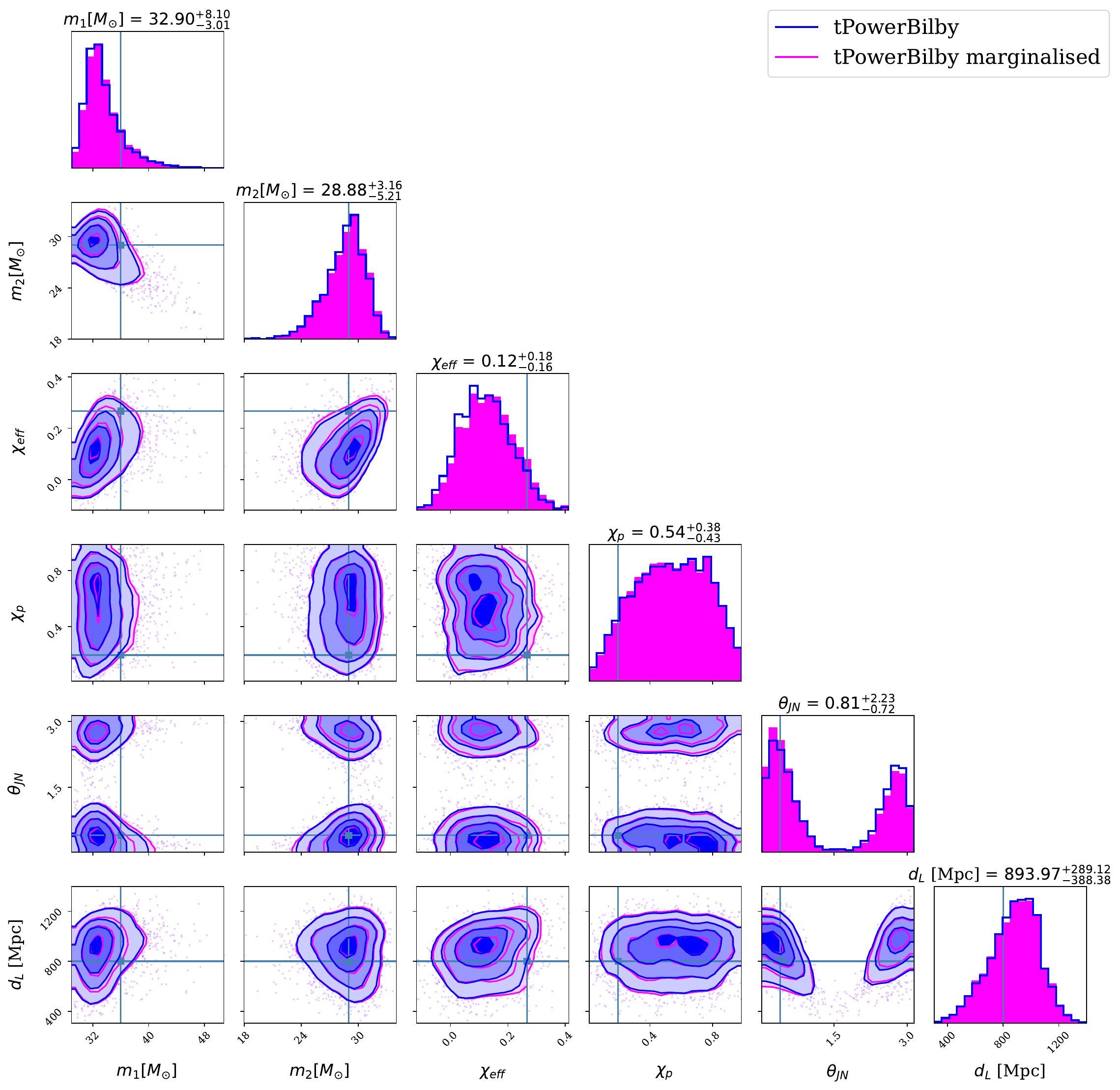}
    \caption{
    Marginalized posterior distributions for the injected signal, with parameters listed in Tab.~\ref{tab:injection_parameters}, showing the masses and spins of the black holes, the inclination angle, and the distance to the binary system. \tpowerbilby{}’s posterior samples are shown in purple, while \tpowerbilby{}’s weighted posterior samples are represented in magenta. The injected values are indicated by vertical and horizontal lines in the respective subplots.}
      
    \label{fig:gw_injection_corner}
\end{figure*}

Figure~\ref{fig:gw_injection_corner} presents the marginalized posterior distributions. The results obtained using the maximum likelihood noise curve from the preferred model are shown in purple, while the results obtained by marginalizing over the noise uncertainties are shown in magenta. The natural logarithm of the Bayes factor for the marginalized model over the unmarginalized model is
$\ln \mathcal{B} \approx 27$ and the efficiency is $\approx 92\%$.
The estimated parameters show good agreement with the injected values, validating \tpowerbilby{}'s noise estimation.  
The very high efficiency indicates that the noise curves used in the marginalization procedure are sufficiently similar to the maximum likelihood noise curve for the purposes of importance sampling.

\begin{table}[h]
    \centering
    \begin{tabular}{|l|c|}
        \hline
        \textbf{Parameter} & \textbf{Value} \\
        \hline
        $m_1$ & 36 $M_{\odot}$\\
        $m_2$ & 29 $M_{\odot}$ \\
        $\chi_{\text{eff}}$ & 0.27 \\
        $\chi_p$ & 0.20\\
        $\text{d}_{\text{L}}$ & 800 Mpc\\
        $\theta_{JN}$  & 0.4 \\
        $\psi$  & 2.7 \\
        Right Ascension  & 1.4 \\
        Declination  & -1.2 \\
        \hline
    \end{tabular}
    \caption{Selected subset of the injected gravitational-wave parameters, including the masses and spins of the black holes, the distance to the binary system, the inclination angle, and the sky location.}
    \label{tab:injection_parameters}
\end{table}

\bibliography{refs}
\end{document}